\documentclass[prd,showpacs,altaffilletter,twocolumn,
superscriptaddress,floatfix,nofootinbib,preprintnumbers]{revtex4-1} 

\pdfoutput=1
\usepackage{amsfonts,amsmath,graphicx,bm} %,siunitx}
\usepackage{float}
\usepackage{dcolumn}
\usepackage[pdftex,
       colorlinks=true,
       urlcolor=blue,       % \href{...}{...} external (URL)
       linkcolor=red,       % \ref{...} and \pageref{...}
       pdfpagemode=None,
       bookmarksopen=true]{hyperref}
\usepackage[capitalise]{cleveref}
\usepackage{ifpdf}
\usepackage{multirow}
\usepackage[colorlinks=true]{hyperref}
\usepackage{url}
\usepackage{slashed}

\newcommand{\glasgow}{SUPA, School of Physics and Astronomy,
  University of Glasgow, Glasgow, G12 8QQ, UK}

\def\today{\number\day\space\ifcase\month\or
January\or February\or March\or April\or May\or June\or
July\or August\or September\or October\or November\or December\fi
\space\number\year}
\newcount\mins \newcount\hours
\def\now{\hours=\time \mins=\time
	\divide\hours by60 \multiply\hours by60 \advance\mins by-\hours
	\divide\hours by60 
	\number\hours:\ifnum\mins<10 0\fi\number\mins }

\newcommand{\RJpsi}{0.2597(27)}       %##
\newcommand{\RATIOEMU}{1.003995(80)}       %##
\newcommand{\gammatau}{2.898(55)\times 10^{-12}\mathrm{GeV}}       %##
\newcommand{\gammamu}{1.116(27)\times 10^{-11}\mathrm{GeV}}       %##
\newcommand{\gammae}{1.120(27)\times 10^{-11}\mathrm{GeV}}       %##

\newcommand{\almtau}{0.5093(42)}       %##
\newcommand{\Fljpsi}{0.4421(55)}       %##
\newcommand{\Afb}{-0.0567(61)}       %##

\newcommand{\gammamufull}{3.00(7)(12)\times 10^{10}\mathrm{s}^{-1}}       %##
\newcommand{\gammamufullGeV}{1.98(5)(8)\times 10^{-14}\mathrm{GeV}}
\newcommand{\Brmufull}{0.0153(4)(6)(3)}       %##
\newcommand{\btoBc}{0.00575(14)(23)(10)(66)}       %##
\begin{document}

\title{Improved Lattice QCD $B_c\to J/\psi$ Vector, Axial-Vector, and Tensor Form Factors}
%% \ShortTitle{Short Title for header}

\author{Judd \surname{Harrison}}
\email[]{judd.harrison@glasgow.ac.uk}
\affiliation{\glasgow}

\collaboration{HPQCD Collaboration}
\email[]{http://www.physics.gla.ac.uk/HPQCD}

\begin{abstract}
We present an update of HPQCD's lattice QCD determination of the $B_c\to J/\psi$ vector and axial-vector form factors, and provide new results for the tensor form factors. We use the Highly Improved Staggered Quark action for all valence quarks, together with the second generation MILC $n_f=2+1+1$ HISQ gluon field configurations. This calculation includes two additional ensembles, one with physically light up and down quarks and $a\approx 0.06 \mathrm{fm}$ and one with $a\approx 0.03\mathrm{fm}$ on which we are able to reach the physical bottom quark mass. Our calculation uses nonperturbatively renormalised current operators and covers the full kinematical range of the decay. We use our recent results for the heavy-charm susceptibilities, as a function of $u=m_c/m_h$, in order to employ the full dispersive parameterisation for $B_c\to J/\psi$ in our physical-continuum extrapolation. We give updated SM predictions $R(J/\psi)=\RJpsi$, $A_{\lambda_\tau}=\almtau$, $F_L^{J/\psi}=\Fljpsi$, and $\mathcal{A}_\mathrm{FB}=\Afb$, reducing uncertainties by $29\%$, $45\%$, $40\%$ and $50\%$ respectively. Since our lattice form factors cover the full kinematic range we can use them to test extrapolations using data in a truncated range, at low-recoil. We investigate different physical continuum parameterisation schemes, with lattice results in the first $1/3$ of the kinematic range near $q^2_\mathrm{max}$. We find that unexpectedly large systematic uncertainties near $q^2=0$ can emerge when extrapolating synthetic data in the high-$q^2$ region if higher order kinematical terms are omitted from the physical continuum extrapolation. This suggests a potentially underestimated systematic uncerainty entering extrapolations of synthetic lattice QCD data for the related $B\to D^*\ell\bar{\nu}$ decay from the high-$q^2$ region into the low-$q^2$ region.
\end{abstract}

%% \tableofcontents

\maketitle

\section{Introduction}

In recent years, lattice QCD studies of semileptonic $B$-meson decays have progressed significantly, with many new results away from zero recoil for decays to both vector and pseudoscalar final states~\cite{EuanBsDs,Cooper:2020wnj,Harrison:2020gvo,Harrison:2021tol,FermilabLattice:2021cdg,Cooper:2021bkt,Colquhoun:2022atw,Parrott:2022rgu,Flynn:2023nhi,Harrison:2023dzh,Aoki:2023qpa}. For $B\to D^{*}\ell\overline{\nu}$ decay, mediated by the Standard Model~(SM) charged-current interaction $b\to cW^-$, the new lattice QCD calculations from Fermilab-MILC~\cite{FermilabLattice:2021cdg}, HPQCD~\cite{Harrison:2023dzh} and JLQCD~\cite{Aoki:2023qpa} have confirmed the long-standing tension between inclusive and exclusive determinations of $V_{cb}$~\cite{FlavourLatticeAveragingGroupFLAG:2024oxs}. While the form factors from these calculations can be fit with reasonable agreement, some tension is seen at the level of $1-2\sigma$ between predictions for observables such as $R(D^*)=\Gamma(B\to D^{*}\tau\overline{\nu}_\tau)/\Gamma(B\to D^{*}\mu\overline{\nu}_\mu)$ made using each calculation~\cite{Martinelli:2023fwm,Bordone:2024weh}. 

The Fermilab-MILC and JLQCD lattice QCD calculations~\cite{FermilabLattice:2021cdg,Aoki:2023qpa} generated data only in the first $\approx 1/3$ and $\approx 1/4$ of the physical range of $w=v_{D^*}\cdot v_B$ respectively. The HPQCD lattice QCD calculation~\cite{Harrison:2023dzh} generated data across the full $w$-range only on the finest ensemble, and with limited statistical precision, resulting in greater uncertainties close to $w_\mathrm{max}$. To overcome limited information in the low-$q^2$ region, a BGL parameterisation~\cite{Boyd:1997kz} is typically used to fit and extrapolate synthetic data points for the form factors across the full kinematical region, so that predictions for integrated observables can be made. A relatively small under-estimation of uncertainties in the low-recoil region, particularly in the slope of the form factors, therefore has the potential to lead to a more significant underestimation of uncertainties in the high-recoil region, and thus in integrated observables such as $R(D^*)$. Moreover, since synthetic form factor data at different values of $w$ typically exhibit large correlations, systematic uncertainties related to higher-order kinematical dependence may be mistakenly deemed insignificant if compared to the total uncertainty. 

The choice of kinematical parameterisation and truncation order used to perform the chiral continuum extrapolation is one potential origin for such underestimated uncertainties. For $B\to D^*\ell\nu$, expansions in powers of $(w-1)$ were used~\cite{FermilabLattice:2021cdg,Colquhoun:2022atw,Harrison:2023dzh}. Fermilab MILC and JLQCD fit the coefficients of $(w-1)^n$ in this expansion directly, including terms up to $(w-1)^2$, while HPQCD used terms up to $(w-1)^{10}$, with coefficients expressed as functions of the continuum BGL expansion parameters, including terms up to $z^4$. HPQCD~\cite{Harrison:2023dzh} also included a correlated fit of the $B_s\to D_s^*$ form factors, related to $B\to D^*$ by chiral parameters including polynomial terms in $M_\pi^2/\Lambda_\chi^2$, as well as the $SU(3)$ rooted-staggered chiral logarithms computed using perturbation theory. This update found differences of order $1-2\sigma$ for $h_{A_2}$ and $h_{A_3}$ relative to the previous $B_s\to D_s^*$ calculation~\cite{Harrison:2021tol}, mostly due to the use of a BGL-like $z$-expansion (omitting the outer functions) to perform the chiral-continuum extrapolation in the earlier work.

$B_c\to J/\psi\ell\nu$ provides an ideal testing ground to investigate the effects of the choice of parameterisation scheme on form factors and observables. The replacement of the light spectator quark with a heavier charm quark relative to the $B\to D^*$ case leads to improved statistical precision, enabling good resolution of the noisier $h_{A_2}$ and $h_{A_3}$ form factors (equivalently $A_{12}$ in the QCD basis), as well as reduced chiral effects from light quark masses. In this work, we provide an update of the HPQCD $B_c\to J/\psi$ form factors~\cite{Harrison:2020gvo}. In addition to methodological improvements, this also work includes the addition of an ensemble with physical light quarks and $a\approx 0.06 \mathrm{fm}$ and an ensemble with heavier-than-physical light quarks and $a\approx 0.03\mathrm{fm}$ on which we are able to simulate the physical $b$ quark with $am_b\approx 0.625$. This calculation includes two ensembles covering the full kinematical range with high statistical precision.

As well as providing an environment to test the methods used to extrapolate to the physical continuum limit, $B_c\to J/\psi\ell\nu$ decays also have the potential to provide important tests of SM lepton flavour universality in the near future, offering insights into the $\approx 3\sigma$ discrepancies seen between SM theory and experiment for $B_{(s)}\to D_{(s)}^{(*)}\ell\nu$ decays. For example, the lepton flavour universality ratio $R(J/\psi)=\Gamma(B_c\to J/\psi\tau\overline{\nu}_\tau)/\Gamma(B_c\to J/\psi\mu\overline{\nu}_\mu)$ has been measured by LHCb~\cite{PhysRevLett.120.121801} using $3\mathrm{fb}^{-1}$ of run 1 data at $7~\mathrm{TeV}$ and $8~\mathrm{TeV}$ centre of mass energies, and a complementary analysis of $R(J/\psi)$ is currently underway at CMS~\cite{CMS:2024uyo}. 

We also calculate the tensor form factors here for the first time. These are defined below in the HQET basis in~\cref{formfactors}. Unlike the SM form factors, $A_{0},~A_{1},~A_{12},$ and $V,$ the tensor form factors only contribute to the differential decay rate when the corresponding new physics couplings in the $b\to c\ell\bar{\nu}$ effective Hamiltonian take nonzero values~\cite{Becirevic:2019tpx}. In~\cite{Harrison:2020nrv} it was demonstrated that shifts consistent with experimental observations away from the SM values of the vector and axial-vector couplings could lead to sizeable shifts in predictions for angular observables and ratios sensitive to lepton flavour universality violation, such as $R(J/\psi)$. With the complete set of SM and tensor form factors, increasingly precise measurements of observables for $B_c\to J/\psi\ell\bar{\nu}$ will allow for improved constraints on new physics couplings complementary to those obtained from $B\to D^*\ell\bar{\nu}$ decay~\cite{Jung:2018lfu,Harrison:2023dzh}. The tensor form factors that we compute directly on the lattice here may also be used to test relations between the tensor and SM form factors derived from nonrelativistic QCD~\cite{Colangelo:2022lpy}.

As in our previous calculation~\cite{Harrison:2020gvo} we use a heavy quark, $h$, with mass $m_h$, in place of the bottom quark, and generate lattice data for $m_h$ ranging from roughly $1.4$ times the charm mass up to the physical bottom quark mass. We denote the $\bar{h}c$ mesons $H_c$. We then fit these results using a physical continuum fit function which describes the $m_h$ dependence, kinematic dependence, and discretisation effects of our data (see~\cref{physcont}) in order to determine the continuum form factors at $m_b$.

The remainder of this paper is organised as follows:
\begin{itemize}
\item In~\cref{sec:lattcalc} we describe the lattice parameters and MILC HISQ 2+1+1 ensembles used in this work. We give details of our correlator fitting approach and current renormalisation, and give the posterior values of the ground-state meson masses.
\item \cref{physcont} contains the details of our physical continuum fit function and extrapolation, based on the dispersive parameterisation for form factors with subthreshold branch cuts given in~\cite{Gubernari:2023puw}. 
\item In~\cref{sec:disc} we provide updated values for $R(J/\psi)$ and other observables sensitive to lepton flavour universality violation. We provide a comprehensive breakdown of form factor uncertainties. We include in this section a comparison to our previous results, as well as the results that would have been obtained from our previous dataset using our updated dispersive parameterisation. A comparison of our tensor form factor results to those derived using NRQCD~\cite{Colangelo:2022lpy} is then shown. We conlude this section by examining our data just in the high-$q^2$ region, and show that omitting higher order kinematical terms from the physical continuum fit in this region results in unexpectedly large discrepancies when extrapolating to the low-$q^2$ region.
\item Finally, in~\cref{sec:conc} we summarise our findings.
\item In~\cref{fitresrecon} we describe in this section how to load our form factor results, including our fully correlated lattice data, which we provide in machine-readable format. In~\cref{corrstabsec,contstabsec} we show tests of the stability of our correlator fitting and continuum extrapolation procedures.
\end{itemize}

\section{Lattice Calculation}
\label{sec:lattcalc}
\begin{table*}
\caption{Details of the MILC $n_f=2+1+1$ HISQ gluon field configurations used in our calculation \cite{PhysRevD.87.054505,PhysRevD.82.074501}. The Wilson flow parameter~\cite{Borsanyi:2012zs}, $w_0$, used to fix the lattice spacing is given in column 3. We use the physical value of $w_0=0.1715(9)\mathrm{fm}$ determined in \cite{PhysRevD.88.074504}. The values of $w_0/a$, 
which are used together with $w_0$ to compute $a$ were taken from \cite{PhysRevD.96.034516,PhysRevD.91.054508,Hatton:2020qhk}. $n_\mathrm{cfg}$ is the number of configurations that we use here and $n_t$ is the number of time sources used on each configuration. $am_{l0}$, $am_{s0}$ and $am_{c0}$ are the masses of the sea up/down, strange and charm quarks in lattice units, while $am_{c}^\mathrm{val}$ and $am_{h}^\mathrm{val}$ are the valence charm and heavy-quark masses respectively.  We also include the approximate mass of the Goldstone pion, computed in~\cite{Bazavov:2017lyh}.\label{tab:gaugeinfo}}
\begin{tabular}{c c c c c c c c c c c c}\hline
 Set &$a$ &$w_0/a$& $L_x\times L_t$ &$am_{l0}$&$am_{s0}$& $am_{c0}$ & $am_{c}^\mathrm{val}$ & $am_{h}^\mathrm{val}$  & $M_\pi$ &$n_\mathrm{cfg}\times n_t $ \\ 
  & $(\mathrm{fm})$& &&& &&&  & $(\mathrm{MeV})$ & \\ \hline
1 & $0.0902$  & 1.9006(20) & $32\times 96 $    &$0.0074$ &$0.037$  & $0.440$       & $0.449$  & $0.65,0.725,0.8$         & $316$ & $1000\times 16$\\
2 & $0.0592$  & 2.896(6)   & $48\times 144  $    &$0.0048$ &$0.024$  & $0.286$     & $0.274$  & $0.427,0.525,0.65,0.8$   & $329$ & $500\times 4$\\
3 & $0.0441$  & 3.892(12)  &$ 64\times 192  $    &$0.00316$ &$0.0158$  & $0.188$   & $0.194$  & $0.5,0.65,0.8$           & $315$ & $375\times 4$\\
4 & $0.0327$  & 5.243(16)  &$ 96\times 288  $    &$0.00223$ &$0.01115$  & $0.1316$ & $0.137$  & $0.4,0.625$              & $309$ & $100\times 4$\\
5 & $0.0879$  & 1.9518(7)  &$ 64\times 96  $    &$0.0012$ &$0.0363$  & $0.432$     & $0.433$  & $0.65,0.725,0.8$   & $129$ & $600\times 8$\\
6 & $0.0568$  & 3.0170(23) & $96\times 192  $    &$0.0008$ &$0.0219$  & $0.2585$   & $0.2585$ & $0.427,0.525,0.65,0.8$   & $135$ & $100\times 4$\\\hline
\end{tabular}
\end{table*}

We generate two- and three-point correlation functions combining heavy ($h$) and charm ($c$) propagators with a variety of $\bar{h}c$, $\bar{c}h$ and $\bar{c}c$ operators on the second generation MILC $n_f=2+1+1$ HISQ gluon field configurations listed in~\cref{tab:gaugeinfo}. The choices of staggered spin-taste operators, valence charm and heavy quark masses, and $J/\psi$ momentum, $\vec{p}'=(ak,ak,0)$, are identical to those used in~\cite{Harrison:2023dzh}. We repeat the spin-taste operators in~\cref{spintastetable} for reference later on. For the additional gluon field ensemble with $a\approx 0.03\mathrm{fm}$ (set 4), we use $am_h^\mathrm{val}=0.4,~0.625$, $am_c^\mathrm{val}=0.1316$ and $ak=0.0,~0.0500,~0.1501,~0.2501$.

\begin{table}
\centering
\caption{Spin-taste operators used in our three-point correlation functions to isolate the form factors. The second column is the operator used for the $H_{c}$, the third for the $J/\psi$ and the fourth column is the operator used at the current. The final column gives the label, referred to in~\cref{fitresrecon}, used to denote the resulting matrix element.\label{spintastetable}}
\begin{tabular}{c | c c c | c }\hline
 &$\mathcal{O}_{H_c}$ & $\mathcal{O}_{J/\psi}$ & $\mathcal{O}_J$ & Label  \\
\hline
$\tilde{J}^{00}_{nn(1,\gamma^{3})}$ & $\gamma_0\gamma_5\otimes \gamma_0\gamma_5$ & $\gamma_1\otimes \gamma_1\gamma_2$ & $\gamma_3\otimes \gamma_3$          & MV \\
$\tilde{J}^{00}_{nn(1,\gamma^{5})}$& $\gamma_5\otimes \gamma_5$ & $\gamma_1\otimes 1$ & $\gamma_5\otimes \gamma_5$                                          & MA0\\
$\tilde{J}^{00}_{nn(3,\gamma^{3}\gamma^{5})}$& $\gamma_5\otimes \gamma_5$ & $\gamma_3\otimes \gamma_3$ & $\gamma_3\gamma_5\otimes \gamma_3\gamma_5$         & MA1\\
$\tilde{J}^{00}_{nn(1,\gamma^{1}\gamma^{5})}$& $\gamma_5\otimes \gamma_5$ & $\gamma_1\otimes \gamma_1$ & $\gamma_1\gamma_5\otimes \gamma_1\gamma_5$         & MA2\\ \hline 
$\tilde{J}^{00}_{nn(3,\sigma^{14})}$ & $\gamma_5\otimes \gamma_5$ & $\gamma_3\otimes \gamma_2\gamma_3$ & $\gamma_0\gamma_1\otimes \gamma_0\gamma_1$         & MT1\\   
$\tilde{J}^{00}_{nn(3,\sigma^{12})}$ & $\gamma_0\gamma_5\otimes\gamma_0 \gamma_5$ & $\gamma_3\otimes \gamma_3$ & $\gamma_1\gamma_2\otimes \gamma_1\gamma_2$ & MT2\\
$\tilde{J}^{00}_{nn(1,\sigma^{23})}$ & $\gamma_0\gamma_5\otimes\gamma_0 \gamma_5$ & $\gamma_1\otimes \gamma_1$ & $\gamma_2\gamma_3\otimes \gamma_2\gamma_3$ & MT23\\ \hline    
\end{tabular}
\end{table}

We fit our lattice correlator data using the standard spectral decomposition for staggered correlation functions, implemented in the \textbf{corrfitter} python package~\cite{corrfitter}:
\begin{align}\label{twopointfit}
\langle  0|&\bar{c}\gamma^\nu c(t) \big(\bar{c}\gamma^\nu c(0)\big)^\dagger| 0 \rangle \nonumber\\
=&\sum_{i}\Big((A^i_n)^2{e^{-tE^{i}_n}}\nonumber-(-1)^{t}(A^i_o)^2{e^{-tE^i_o}}\Big),\\
\langle  0|&\big(\bar{h}\gamma^5 c(t)\big)^\dagger\bar{h}\gamma^5 c(0) | 0 \rangle \nonumber\\
=&\sum_{i}\Big((B^i_n)^2{e^{-tM^{i}_n}}-(-1)^{t}(B^i_o)^2{e^{-tM^i_o}}\Big)
\end{align}
and
\begin{align}\label{threepointfit}
\langle  0|\bar{c}\gamma^\nu c(T) ~ \bar{c}\Gamma h(t)& ~ \bar{h}\gamma^5 c(0)| 0 \rangle \nonumber\\
=\sum_{i,j}\Big(&{  A^i_n B^j_n J^{ij}_{nn(\nu,\Gamma)} e^{-(T-t)E^{i}_n - tM^{j}_n} }\nonumber\\
+{(-1)^{T+t}}  &A^i_o B^j_n J^{ij}_{on(\nu,\Gamma)} e^{-(T-t)E^i_o - tM^{j}_n} \nonumber\\
+{(-1)^{t}}  &A^i_n B^j_o J^{ij}_{no(\nu,\Gamma)} e^{-(T-t)E^{i}_n - tM^j_o} \nonumber\\
+{(-1)^{T}}  &A^i_o B^j_o J^{ij}_{oo(\nu,\Gamma)} e^{-(T-t)E^i_o - tM^j_o} \Big).
\end{align}
Here, and for the remainder of this section, we write the meson interpolating operators in terms of the naive quark fields rather than the equivalent staggered fields and spin-taste operators for simplicity. We have also left implicit the sums over spatial dimensions for each fermion bilinear.

The ground state parameters of these fits are related straightforwardly to the energies and matrix elements we require for computing form factors, giving
\begin{equation}
J^{00}_{nn(\nu,\Gamma)} = \sum_{\lambda}\frac{\epsilon^\nu(p',\lambda) \langle  J/\psi(p',\lambda ) |\bar{c}\Gamma h |H_c(p)\rangle}{\sqrt{2E_{J/\psi}2M_{H_c}\left(1+{\vec{p}}_{\nu}^{~\prime 2}/M_{J/\psi}^2\right)}}\label{relnorm}
\end{equation}
where ${\vec{p}}~'_{\nu}$ is the $\nu$ component of the $J/\psi$ spatial momentum, with $\nu$ corresponding to the Lorentz index of the $J/\psi$ vector interpolating operator. The matrix elements are related to the form factors, $h_X$, in the HQET basis by
\begin{align}\label{formfactors}
\frac{\langle J/\psi|\bar{c}\gamma^5 h|\overline{{H_c}}\rangle}{\sqrt{M_{H_c}M_{J/\psi}}}             =& -(\epsilon^*\cdot v)h_P,\nonumber\\
\frac{\langle J/\psi|\bar{c}\gamma^\mu h|\overline{{H_c}}\rangle}{\sqrt{M_{H_c}M_{J/\psi}}}           =&~ i\varepsilon^{\mu\nu\alpha\beta}\epsilon^{*}_\nu v^\prime_\alpha v_\beta h_V,\nonumber\\
\frac{\langle J/\psi|\bar{c} \gamma^\mu \gamma^5 h|\overline{{H_c}}\rangle}{\sqrt{M_{H_c}M_{J/\psi}}} =& ~\big[ h_{A_1}(w+1)\epsilon^{*\mu}\nonumber\\
-h_{A_2}&(\epsilon^*\cdot v)v^\mu-h_{A_3}(\epsilon^*\cdot v)v^{\prime\mu} \big],\nonumber\\
\frac{\langle J/\psi|\bar{c}\sigma^{\mu\nu} h|\overline{{H_c}}\rangle}{\sqrt{M_{H_c}M_{J/\psi}}}      =& -\varepsilon^{\mu\nu\alpha\beta}\big[ h_{T_1}\epsilon^*_\alpha(v+v^\prime)_\beta \nonumber\\
+h_{T_2}&\epsilon^*_\alpha(v-v^\prime)_\beta +h_{T_3}(\epsilon^*\cdot v)v_\alpha v^\prime_\beta \big].
\end{align}

\subsection{Current Renormalisation}
\label{sec:renorm}
\begin{table}
\centering
\caption{$Z_V(\mu=2\mathrm{GeV})$ at zero valence quark mass from~\cite{Hatton:2019gha} and~\cite{Hatton:2020qhk} in the RI-SMOM scheme and $Z_T(\mu=4.8~\mathrm{GeV})$ from~\cite{Hatton:2020vzp} for the tensor operators. $Z_V$ on sets 5 and 6 are equal to those on sets 1 and 2 respectively. On set 4, we use a value of $Z_T$ obtained by extrapolating the other values in $a^2$ as described in the text. \label{tab:Z}}
\begin{tabular}{c c c }\hline
 Set & $Z_V(\mu=2\mathrm{GeV})$  & $Z_T(\mu=4.8~\mathrm{GeV})$ \\ \hline
1 & $0.98445(11)$&$1.0029(43)$  \\\hline
2 &  $0.99090(36)$&$1.0342(43)$   \\\hline
3 &  $0.99203(108)$&$1.0476(42)$  \\\hline
4 &  $0.99296(21)$&$1.0570(50)$ \\ \hline
5 &  $0.98445(11)$&$1.0029(43)$ \\ \hline
6 &  $0.99090(36)$&$1.0342(43)$   \\\hline
\end{tabular}
\end{table} 

The lattice operators we use must be related to their continuum counterparts by renormalisation. Since we determine the pseudoscalar form factor via the partially conserved axial current relation, no renormalisation factor is required. For the vector current, we use the renormalisation factors, $Z_V$, computed in the RI-SMOM scheme in~\cite{Hatton:2019gha} and~\cite{Hatton:2020qhk}. We also use these for the axial-vector current, using the chiral symmetry of the HISQ action~\cite{Sharpe:1993ur} together with the absence of condensate contamination in $Z_V$~\cite{Hatton:2019gha}. The tensor renormalisation factors, $Z_T$, were computed in~\cite{Hatton:2020vzp} using an intermediate RI-SMOM scheme, matched to $\overline{\mathrm{MS}}$ at a scale $\mu=2\mathrm{GeV}$. We run these to $\mu=\overline{m}_h(\overline{m}_h)$ using the 3-loop anomalous dimension~\cite{Gracey:2000am} before multiplying our tensor matrix elements. For $\overline{m}_h(\overline{m}_h)$, we first compute $\overline{m}_h(3\mathrm{GeV})=\overline{m}_c(3\mathrm{GeV})/(am_c/am_h)$ from our charm and heavy quark masses together with the value of $\overline{m}_c(3\mathrm{GeV})=0.9858(51)\mathrm{GeV}$ from~\cite{Hatton:2020qhk}, which we then run to $\overline{m}_h(\overline{m}_h)$ using $\alpha_{\overline{\mathrm{MS}}}(5\mathrm{GeV},n_f=4)=0.2128(25)$ from~\cite{PhysRevD.91.054508} together with the 4-loop running~\cite{vanRitbergen:1997va}. Following~\cite{Harrison:2024iad} we extrapolate the values of $Z_T$ from~\cite{Hatton:2020qhk} as a function of $a^2$ to set 4 using the simple function
\begin{align}
Z_T(a,\mu=2\mathrm{GeV}) =  \sum_{i=0}^{i=4}\left(c_i + \sum_{j=1}^{j=3} b_{ij}\left(\frac{a\mu}{\pi}\right)^{2j}\right)\alpha_s(\pi/a)^i
\end{align}
with Gaussian priors of $0\pm 2$ for the coefficients $c_i$ and $b_{ij}$. Varying $\mu$ or $\pi/a$ by $\pm 50\%$ or increasing the maximum order that we sum to in $i$ or $j$ has does not change the extrapolated value of $Z_T$ significantly. We neglect correlations between the renormalisation factors, which have only very small statistical uncertainties, and our lattice data. The values of $Z_V$ and $Z_T$ that we use are given in~\cref{tab:Z}. For simplicity, we will leave the renormalisation factors implicit in the remaining discussion.

\subsection{Correlator Fitting}

\begin{figure}
\centering                            
\includegraphics[width=0.5\textwidth]{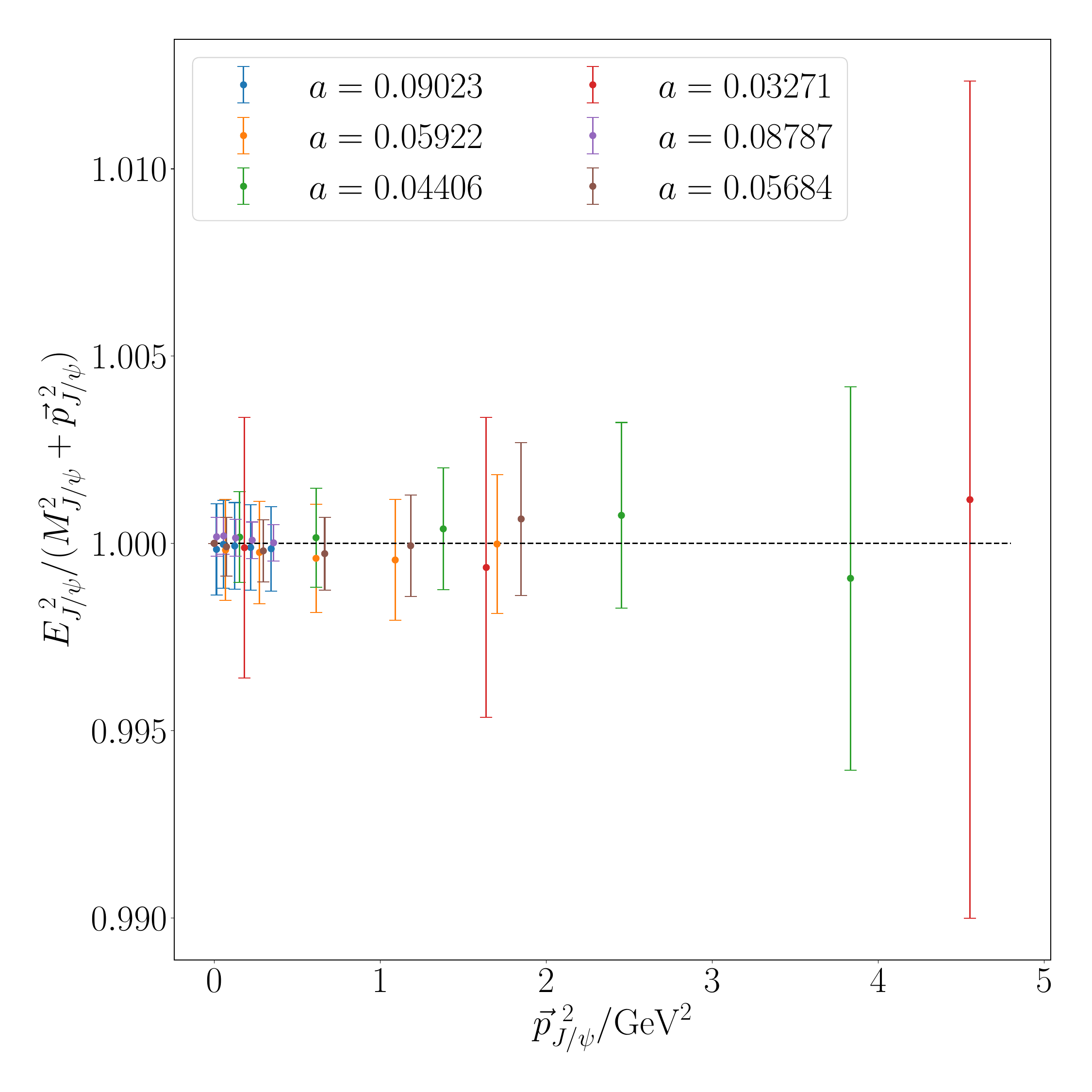}
\caption{\label{speedoflight} Ratios of lattice energies, $E_{J/\psi}$, extracted from correlator fits to the energy computed from the relativistic dispersion relation, $aE_{J/\psi}=\sqrt{aM_{J/\psi}^2+(a\vec{p}')^2}$. We see that our data is in good agreement with this relation across the full range of momenta considered and to high precision.}
\end{figure}

\begin{table}
\centering
\caption{Correlator fit priors. We take $\Delta E^{(o)}_i=\Lambda_\mathrm{QCD}\times 1.0(0.75)$ where $\Delta E^{(o)}_i = E^{(o)}_{i+1}-E^{(o)}_{i},~i\geq 0$ and here for our correlator fits we take $\Lambda_\mathrm{QCD}=0.75~\mathrm{GeV}$. In the table we have defined $\Omega_{H_{c}}=M_{H}^\mathrm{max}+m_h-0.8$ (except on set 4, where we use $\Omega_{H_{c}}=(M_B^\mathrm{phys}+m_h-0.625+m_c)$) and $\Omega_{J/\psi}=\sqrt{M_{J/\psi}^2+2k^2}$, following the relativistic dispersion relation.\label{priortable}}
\begin{tabular}{ c c c c c }\hline
 Prior & $J/\psi(k)$ & $H_c$\\\hline
$E_n^0/\mathrm{GeV}$	 	&$\Omega_{J/\psi}\times 1.0(0.3)$ 	&$\Omega_{H_{c}}\times 1(0.3)$ 	\\
$E^0_o/\mathrm{GeV}$		&$\Omega_{J/\psi}\times 1.2(0.5)$	&$\Omega_{H_{c}}\times 1.2(0.5)$		\\
$A(B)^{n(o)}_i$			&$0.1(5.0)$			&$0.1(5.0)$\\\hline
\end{tabular}
\end{table}

\begin{table}
\centering
\caption{Details of correlator fit parameters. $\Delta T$ indicates the number of data points excluded from the extremities of correlation functions, $n_\mathrm{exp}$ is the number of non-oscillating and time-oscillating exponentials included in our correlator fits to~\cref{twopointfit,threepointfit}. \label{fitparams} Following~\cite{Harrison:2023dzh} we estimate $\chi^2/\mathrm{dof}$ by introducing SVD and prior noise, implemented in~\cite{corrfitter}. The resulting value of $\chi^2/\mathrm{dof}$ is given in column 7. We use the fit parameters in bold for our subsequent analysis. $\delta$ labels variations on our base fit used to investigate the stability of our analysis.}
\begin{tabular}{ c c c c c c c c }\hline
Set & $n_\mathrm{exp}$ & $\Delta T_\mathrm{3pt}$ & $\Delta T_\mathrm{2pt}^{J/\psi}$& $\Delta T^{H_{c}}_\mathrm{2pt}$ & SVD cut & $\chi^2/\mathrm{dof}$ & $\delta$ \\ \hline
1  & \textbf{3} & \textbf{4} & \textbf{8} & \textbf{8} & \textbf{0.005} & \textbf{1.07} & \\
   & {3} & {5} & {10} & {10} & {0.005} & {1.02} & 1\\
   & {3} & {3} & {7} & {7} & {0.02} & {1.00} & 2\\\hline
2  & \textbf{3} & \textbf{4} & \textbf{9} & \textbf{9} & \textbf{0.005} & \textbf{1.03}& \\
   & {3} & {6} & {12} & {12} & {0.005} & {1.05} & 1\\
   & {3} & {4} & {9} & {9} & {0.0075} & {1.03} & 2\\\hline
3  & \textbf{3} & \textbf{6} & \textbf{12} & \textbf{12} & \textbf{0.005} & \textbf{1.04}& \\
   & {3} & {7} & {14} & {14} & {0.005} & {1.04} & 1\\
   & {3} & {6} & {12} & {12} & {0.01} & {1.01} & 2\\\hline
4  & \textbf{3} &  \textbf{11} & \textbf{22} & \textbf{22} & \textbf{0.01} & \textbf{0.90}& \\
   & {3} & {14} & {28} & {28} & {0.01} & {0.97} & 1\\
   & {3} & {11} & {22} & {22} & {0.05} & {0.93} & 2\\\hline
5  & \textbf{3} &  \textbf{3} & \textbf{7} & \textbf{7} & \textbf{0.005} & \textbf{1.06}& \\
   & {3} & {5} & {10} & {10} & {0.005} & {1.05} & 1\\
   & {3} & {3} & {7} & {7} & {0.01} & {1.06} & 2\\\hline
6  & \textbf{3} &  \textbf{3} & \textbf{7} & \textbf{7} & \textbf{0.02} & \textbf{1.01}& \\
   & {3} & {5} & {10} & {10} & {0.02} & {1.05} & 1\\
   & {3} & {3} & {7} & {7} & {0.0075} & {1.03} & 2\\\hline
\end{tabular}
\end{table}

\begin{table}
\centering
\caption{Fit results for the $J/\psi$ masses for the local spin-taste operator $\gamma_1\otimes \gamma_1$ and $1-$link operators $\gamma_1\otimes 1$ and $\gamma_1\otimes \gamma_1\gamma_2$ used in our calculation. \label{charmMasses}}
\begin{tabular}{ c c c c }
\hline
Set & $\gamma_1\otimes \gamma_1$&  $\gamma_1\otimes 1$ & $\gamma_1\otimes \gamma_1\gamma_2$  \\\hline
1&1.41397(62)	&1.41445(79)	&1.41418(66)	\\
\hline
2&0.92974(44)	&0.92978(59)	&0.92988(57)	\\
\hline
3&0.69221(30)	&0.69225(40)	&0.69227(39)	\\
\hline
4&0.51386(73)	&0.5139(11)	&0.51378(92)	\\
\hline
5&1.37849(28)	&1.37869(55)	&1.37842(44)	\\
\hline
6&0.89224(25)	&0.89231(37)	&0.89227(33)	\\
\hline
\end{tabular}
\end{table}

\begin{table}
\centering
\caption{Fit results for the $H_c$ masses for the local spin-taste operators $\gamma_5\otimes \gamma_5$ and $\gamma_0\gamma_5\otimes \gamma_0\gamma_5$ that we use in our calculation. \label{Massesheavy}}
\begin{tabular}{ c c c c }
\hline
Set & $am_h$ & $\gamma_5\otimes \gamma_5$& $\gamma_0\gamma_5\otimes \gamma_0\gamma_5$ \\\hline
1&	0.65	&1.57248(24)	&1.57384(48)	\\
&	0.725	&1.64718(26)	&1.64864(49)	\\
&	0.8	&1.72014(36)	&1.72190(50)	\\
\hline
2&	0.427	&1.06721(20)	&1.06748(25)	\\
&	0.525	&1.17254(20)	&1.17283(25)	\\
&	0.65	&1.30310(20)	&1.30341(25)	\\
&	0.8	&1.45416(20)	&1.45448(25)	\\
\hline
3&	0.5	&1.01190(13)	&1.01201(15)	\\
&	0.65	&1.16998(13)	&1.17011(15)	\\
&	0.8	&1.32186(13)	&1.32199(15)	\\
\hline
4&	0.4	&0.80134(24)	&0.80153(25)	\\
&	0.625	&1.04277(26)	&1.04297(29)	\\
\hline
5&	0.65	&1.55435(13)	&1.55594(20)	\\
&	0.725	&1.62927(14)	&1.63092(20)	\\
&	0.8	&1.70263(14)	&1.70434(21)	\\
\hline
6&	0.427	&1.04807(15)	&1.04832(17)	\\
&	0.525	&1.15366(15)	&1.15392(17)	\\
&	0.65	&1.28448(15)	&1.28476(17)	\\
&	0.8	&1.43577(15)	&1.43607(17)	\\
\hline
\end{tabular}
\end{table}

We fit all correlation functions, defined in~\cref{twopointfit,threepointfit}, on each ensemble simultaneously. For ground-state priors we take, in lattice units, $E^{J/\psi}_0=\sqrt{M_{J/\psi}^2+2k^2}\times 1(0.3)~\mathrm{GeV}$ for the $J/\psi$ energies and $M_{H_c}^0=(M_{H}^\mathrm{max}+m_h-0.8+m_c)\times 1(0.3)~\mathrm{GeV}$ for the $H_c$ masses, with $M_{H}^\mathrm{max}$ the approximate value of $M_{H}$ from \cite{Harrison:2023dzh} for the largest value of $am_h=0.8$, corresponding to $2.86\mathrm{GeV}$, $3.9\mathrm{GeV}$ and $5.0\mathrm{GeV}$ on sets 1 and 5, 2 and 6, and 3 respectively. For set 4, where the maximum value of $am_h$ is 0.625 we use $M_{H_c}^0=(M_B^\mathrm{phys}+m_h-0.625+m_c)$. Our priors for the lowest oscillating state energies, as well as amplitudes, are given in~\cref{priortable}. For the matrix elements, $J^{ij}_{n(o)n(o)}$, we take priors $0(1)$ for all except those proportional to $ak$. For these, following~\cite{Harrison:2023dzh}, we divide by $ak$ before fitting. We increase the uncertainty on the corresponding priors for the oscillating state matrix elements $J^{ij}_{no}$, $J^{ij}_{oo}$, and $J^{ij}_{oo}$ by a factor of 4 relative to $J^{ij}_{nn}$ to account for this rescaling, and take priors of $0(4)$.

We truncate our correlator fits at 3 exponentials. In order to suppress excited state contamination, we also exclude correlator data points close to the minimum and maximum values of $t$ in our correlation functions. Our correlator fits require the use of an SVD cut~(see Appendix D of~\cite{Dowdall:2019bea}) due to the large number of correlator data points relative to the number of configurations making up each ensemble. In order to estimate the quality of our fits, we compute $\chi^2/\mathrm{dof}$ with SVD and prior noise. Unlike~\cite{Harrison:2023dzh}, we bin all time sources on each configuration for sets 1, 2, 3, and 5. It is only on sets 4 and 6, with 100 configurations, that we do not bin over time sources. The number of data points excluded from the ends of our correlators, which we denote $\Delta T$, as well as the SVD cut used, are given in~\cref{fitparams}. In~\cref{fitparams} we also give the parameters used in our fit variations that we use in~\cref{corrstabsec} to illustrate that our results are insensitive to increasing $\Delta T$ or increasing the size of the SVD cut.

The ground state energies for the $J/\psi$ and $H_c$ that we extract from our fits are given in~\cref{charmMasses,Massesheavy}. Note that these agree well with the values given in~\cite{Harrison:2020gvo}, though have somewhat larger uncertainties owing to the larger SVD cut required by the inclusion of additional tensor data here. We use the masses from the local $\gamma_1\otimes \gamma_1$ $J/\psi$ operators and the $\gamma_5\otimes \gamma_5$ $H_c$ operators when converting between matrix elements and form factors and in our physical continuum extrapolation discussed below. Since the taste splittings are very small (less than $0.1\%$ for any of the meson masses considered here) and amount to discretisation effects which vanish in the continuum limit, this choice is not significant. We use $J/\psi$ energies computed from the relativistic dispersion relation, $aE_{J/\psi}=\sqrt{aM_{J/\psi}^2+(a\vec{p}')^2}$, since the momentum is known exactly. We plot the speed of light computed on each ensemble for the range of momenta used in this calculation in~\cref{speedoflight}, where we see that all of the energies we extract agree with the dispersion relation within $0.25\%$ and within uncertainties.

The effect of nonequilibrated topological charge on the finest lattices was discussed in~\cite{Bazavov:2017lyh} and~\cite{Bernard:2017npd}, where the leading order corrections for heavy-light masses and decay constants were estimated using chiral perturbation theory and found to be very small. These effects were shown to decrease with light quark mass, with the heavy-strange decay constant corrections a factor of $(m_l/m_s)^2=1/25$ smaller than the up/down case. These effects are expected to be even smaller for the charm quark case, where the charm quark also decouples from the chiral theory. As such, we neglect the effects of the poorly sampled distribution of topological charge on sets 3 and 4, which are expected to be much smaller than our statistical uncertainties.

The ground state matrix elements, defined in~\cref{relnorm}, are provided in the supplementary material in machine-readable format. The values given include the renormalisation factors in~\cref{tab:Z}, and instructions for their use are given in~\cref{fitresrecon}.

\section{Physical Continuum Extrapolation}
\label{physcont}

%\begin{figure*}
%\centering                            
%\includegraphics[width=1\textwidth]{figures/BcJpsi_ALL_z_no_a2mh_correction.pdf}
%\caption{\label{ffplots}Lattice data for all SM and tensor form factors together with the result of our physical continuum extrapolation described above, shown as the blue line %and error band. The red dashed lines drawn for the SM form factors correspond to the $\pm 1\sigma$ confidence interval of our previous result~\cite{Harrison:2020gvo}.}
%\end{figure*}

\begin{figure*}
\centering                            
\includegraphics[width=1\textwidth]{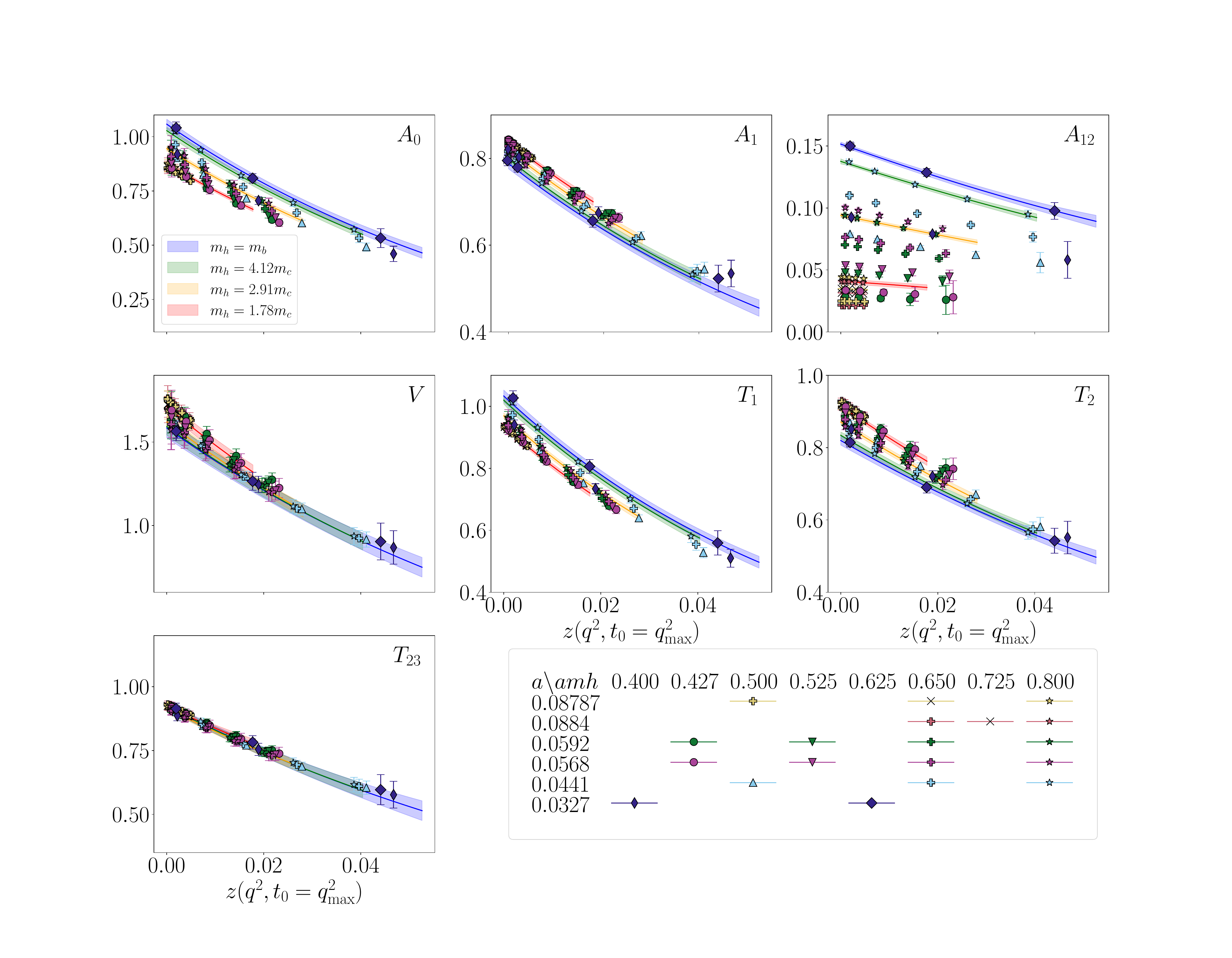}
\caption{\label{ffcorrectedplots}Lattice data for all SM and tensor form factors for $B_c\to J/\psi$ together with the result of our continuum extrapolation described above, for several heavy quark masses. The physical result at $m_h=m_b$ is shown as the blue line and error band, while results for $m_h =4.12m_c,~2.91m_c$ and $1.78m_c$ are shown in green, orange and red respectively. The blue $m_h=m_b$ band is plotted for the physical $q^2$ range, while the unphysical $m_h=4.12m_c,~2.91m_c$ and $1.78m_c$ bands have been been plotted for $q^2$ ranges chosen to aid clarity, beginning at $q^2_\mathrm{max}$. Note that to determine $M_{H_c}$ for these unphysical masses we fit our lattice masses with a simple function of $\overline{m}_h(\overline{m}_h)$ including $am_c$ and $am_h$ discretisation effects, as well as scale dependence through $\alpha(\overline{m}_h)$. The data points shown here have been corrected for mistuned charm and sea quark masses and discretisation effects using the fit posteriors, using~\cref{correction}.}
\end{figure*}
\begin{figure*}
\centering                            
\includegraphics[width=1\textwidth]{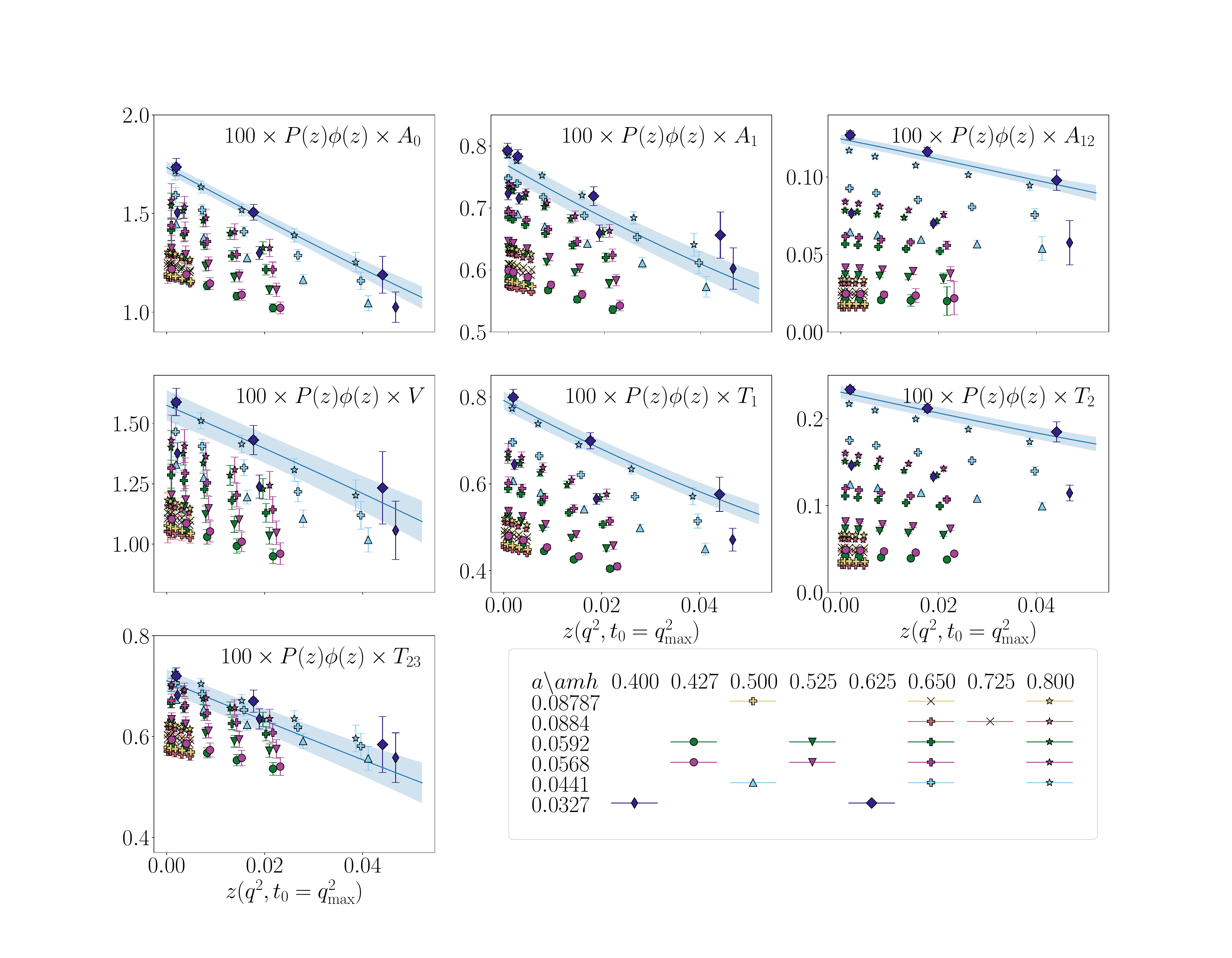}
\caption{\label{ffPphiplots}Lattice data for all SM and tensor form factors for $B_c\to J/\psi$ together with the result of our physical continuum extrapolation described above, shown as the blue line and error band. The data points and continuum results shown here have been multiplied by $P^Y(z,t_+,t_0)\phi^Y(z,t_+,t_0)$.}
\end{figure*}

\begin{figure*}
\centering                            
\includegraphics[width=1\textwidth]{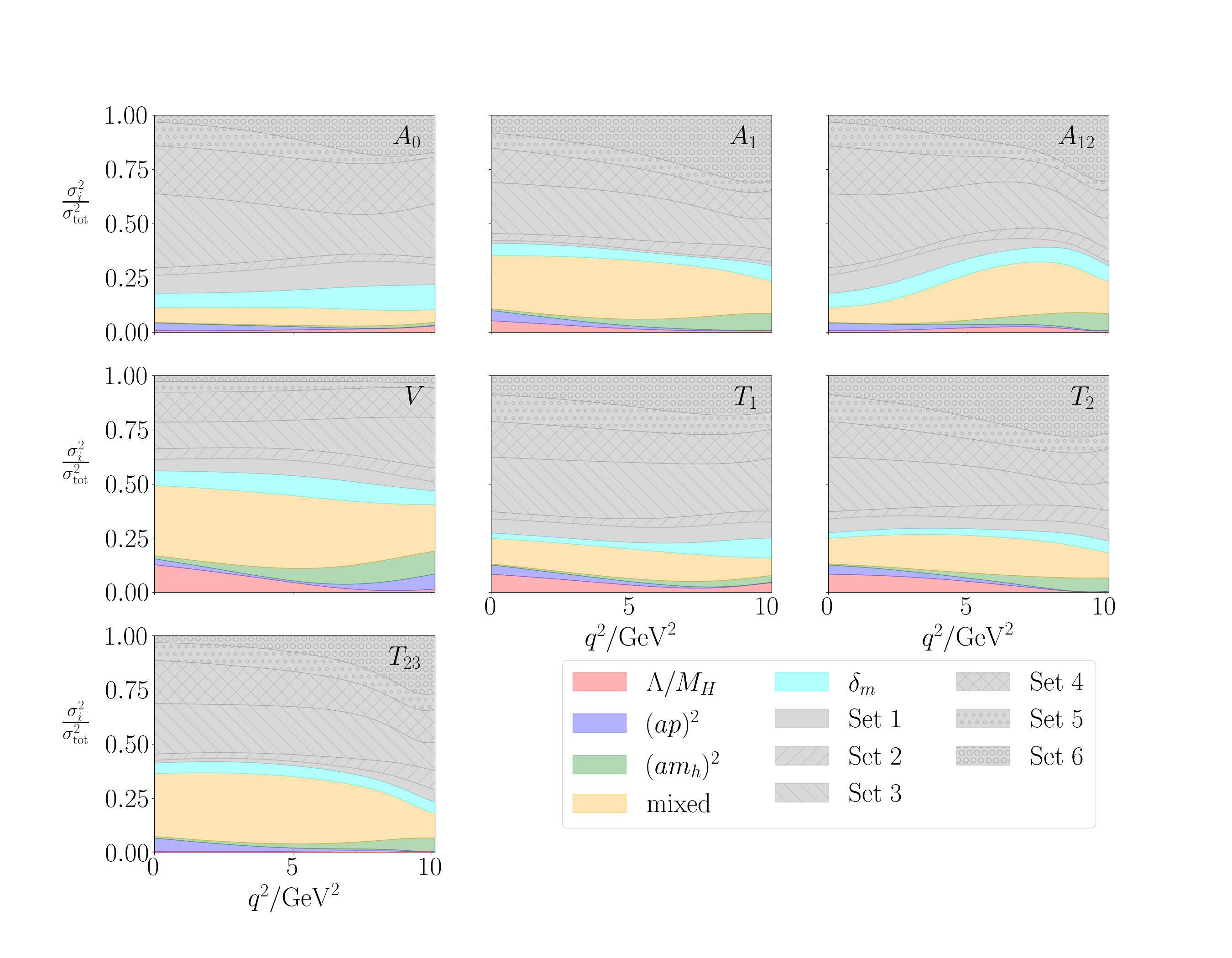}
\caption{\label{fferrorbandplots} Here we show the partial variance of each form factor broken down into contributions from $\Lambda_\mathrm{QCD}/M_H$, $(am_h)^2$, $(ap)^2$, and quark mass mistuning effects, as well as mixed terms combining these, and the statistical uncertainty of our data. Unlike~\cite{Harrison:2023dzh}, we are able to resolve the leading order $\Lambda_\mathrm{QCD}/M_H$ and $(am_h)^2$ uncertainties well enough that the mixed uncertainties (which now include terms $\propto (a\Lambda_\mathrm{QCD}/\pi)^2$) contribute the dominant uncertainties.}
\end{figure*}

The dispersive parameterisation is formulated in the QCD basis~\cite{Gubernari:2023puw}, which is related to the HQET basis given in~\cref{formfactors} by~\cite{Bordone:2019vic}
\begin{align}\label{formfactorsQCDSM}
V   &= h_V\frac{1+r}{2 \sqrt{r}},\nonumber\\
A_0 &= \frac{1}{2 \sqrt{r}}\left(h_{A_1}(1+w)+h_{A_2}(rw-1)+h_{A_3}(r-w)\right),\nonumber\\
A_1 &= \frac{\sqrt{r}}{1+r}h_{A_1}  (1+w),\nonumber\\
A_2 &= \frac{1+r}{2 \sqrt{r}}\left(h_{A_2} r+h_{A_3}\right),
\end{align}
for the SM form factors and by
\begin{align}\label{formfactorsQCDNP}
T_1 &= -\frac{1}{2\sqrt{r}}\left(  (1-r)h_{T_2} - (1+r)h_{T_1}  \right),\nonumber\\
T_2 &= \frac{1}{2\sqrt{r}}\left( \frac{2r(w+1)h_{T_1}}{1+r} - \frac{2r(w-1)h_{T_2}}{1-r}  \right)\nonumber\\
T_3 &= \frac{1}{2\sqrt{r}}\big(  (1-r)h_{T_1} - (1+r)h_{T_2} + (1-r^2)h_{T_3}  \big)
\end{align}
for the tensor form factors. It is conventional to define the related definite-helicity form factors
\begin{align}\label{helicityffsA12T23}
    A_{12} =& \frac{(M_{H_c} + M_{J/\psi})^2 (M_{H_c}^2 - M_{J/\psi}^2 - q^2)  A_1 - \lambda A_2}{16M_{H_c} M_{J/\psi}^2 (M_{H_c} + M_{J/\psi})}, \nonumber\\
    T_{23} =& \frac{(M_{H_c}^2 - M_{J/\psi}^2) (M_{H_c}^2 + 3 M_{J/\psi}^2 - q^2)  T_2 -  \lambda T_3}{8 M_{H_c} M_{J/\psi}^2 (M_{H_c} - M_{J/\psi})},
\end{align}
which diagonalise the bounds. Here $\lambda=(M_{H_c}^2 - M_{J/\psi}^2 -q^2)^2-4M_{J/\psi}^2q^2$.

Our fit function is chosen to give the dispersive parameterisation described in~\cite{Gubernari:2023puw} in the continuum limit, $a\to 0$. Rather than powers of $z(q^2,t_+,t_0)=\frac{\sqrt{t_+-q^2}-\sqrt{t_+-t_0}}{\sqrt{t_+-q^2}+\sqrt{t_+-t_0}}$, the dispersive parameterisation for form factors with subthreshold branch cuts is constructed from orthonormal polynomials on the unit circle, 
\begin{align}\label{orthopols}
\int_{-\alpha(t_X)}^{\alpha(t_X)} d\theta p_n(e^{i\theta})p_m(e^{-i\theta}) = \delta_{nm},
\end{align} 
where $\alpha(t_X) = \arg z(t_X,t_+,t_0)$, with $t_+=(M_H+M_{D^*})^2$ the two particle production threshold for the $\bar{c}\Gamma h$ current and $t_X=(M_{H_c}+M_{J/\psi})^2$ the start of the physical region for $H_cJ/\psi$ production. Each polynomial, $p_n(z)$, includes only terms $z^{ 0\leq m\leq n}$, and they reduce to $p_n(z)=z^n/\sqrt{2\pi}$ when $t_X=t_+$ and hence $\alpha(t_X)=\pi$. We have checked the orthonormal polynomials which we implement up to $\mathcal{O}(z^5)$ against the values given in the \texttt{EOS} software~\cite{EOSAuthors:2021xpv} version 1.0.13~\cite{EOS:v1.0.13}.

The continuum parameterisation is then given by
\begin{align}\label{eq:ff}
F^Y(z) = &\frac{1}{P^Y(z,t_+,t_0)\phi^Y(z,t_+,t_0)}\sum_{n} a_n^Y p_n(z) \mathcal{N}_n.
\end{align}
Here, $P^Y(z(q^2,t_+,t_0)) = \prod_{i}z(q^2,t_+,M_{\mathrm{pole},i}^2)$ are Blaschke factors which depend on the masses, $M_{\mathrm{pole},i}$, of single particle $\bar{h}c$ states created by the current below the pair production threshold $t_+$. $\phi(z,t_+,t_0)$ are outer functions, defined in \cite{Gubernari:2023puw}, which are analytic on the open unit disk in $z$ and also depend on the susceptibilities, $\chi$. The susceptibilities for $\bar{b}c$ currents may be computed using either perturbation theory~\cite{Hoff:2011ge,Grigo:2012ji,Bharucha:2010im} or lattice QCD~\cite{DiCarlo:2021dzg}. Here, we use our recent lattice QCD calculation~\cite{Harrison:2024iad} of the (pseudo)scalar, (axial-)vector and (axial-)tensor susceptibilities as a function of $u=m_c/m_h$. Agreement of the nonperturbative lattice QCD susceptibilities with expectations from perturbation theory was found to be good, except for the (axial-)tensor case~\cite{Harrison:2024iad}. The coefficients, $a_n^Y$, satisfy the bounds $\sum_n |a_n^Y|^2\leq 1$, where there is a bound for each current, and the sum over $Y$ runs over form factors which contribute in each case. In our continuum fit function,~\cref{eq:ff}, we include polynomials up to and including $\mathcal{O}(z^4)$. We also include the quark mass mistuning term,
\begin{equation}\label{mistuning}
\mathcal{N}^{Y^{(s)}}_n = 1 + A^{Y}_n \delta_{m_c}^\mathrm{val}+ B^{Y}_n \delta_{m_c}^\mathrm{sea}+ C^{Y}_n \delta_{m_s}^\mathrm{sea}+D^{Y}_n \delta_{\chi}
\end{equation}
where
\begin{align}
\delta_{m_c}^\mathrm{val} &= (am_c^\mathrm{val}-am_c^\mathrm{tuned})/am_c^\mathrm{tuned},\nonumber\\
\delta_{m_c}^\mathrm{sea} &= (am_{c}^\mathrm{sea}-am_c^\mathrm{tuned})/am_c^\mathrm{tuned},\nonumber\\
\delta_{m_{s}}^\mathrm{sea} &= (am_{s}^\mathrm{sea} - am_{s}^\mathrm{tuned})/(10am_{s}^\mathrm{tuned}),\nonumber\\
\delta_\chi&=\left(\frac{M_{\pi}}{\Lambda_\chi}\right)^2-\left(\frac{M_{\pi}^\mathrm{phys}}{\Lambda_\chi}\right)^2.
\end{align}
In the last equation we use the pion masses given in~\cite{Bazavov:2017lyh}. We take Gaussian priors of $0(1)$, $0(0.1)$, $0(0.2)$, and $0(0.2)$ for $A^{Y}_n$, $B^{Y}_n$, $C^{Y}_n$, and $D^{Y}_n$ respectively, motivated by the 1-loop nature of sea quark mass mistuning effects together with the findings in~\cite{PhysRevD.91.054508} for the size of sea charm quark mass mistuning effects. The tuned values of the charm and strange quark masses are given by
\begin{equation}\label{amctuned}
am_c^\mathrm{tuned} = am_c^\mathrm{val}\left(\frac{M_{J/\psi}^\mathrm{phys}}{M_{J/\psi}}\right)^{1.5},
\end{equation}
and
\begin{equation}
am_s^\mathrm{tuned} = am_s^\mathrm{val}\left(\frac{M_{\eta_s}^\mathrm{phys}}{M_{\eta_s}}\right)^2.
\end{equation}
Here we take $M_{\eta_s}^\mathrm{phys}=0.6885(22)\mathrm{GeV}$ determined from the masses of the pion and kaon in~\cite{PhysRevD.88.074504}. For sets 1, 2, 3, and 5 we use the values of $M_{\eta_s}$ and $am_s^\mathrm{val}$ given in~\cite{EuanBsDsstar}. On sets 4 and 6 we use $aM_{\eta_s}=0.11430(50)$ and $aM_{\eta_s}=0.198239(76)$ determined from fits to two-point correlation functions with valence strange masses of $am_s^\mathrm{val}=0.0117$ and $am_s^\mathrm{val}=0.0219$ respectively. The power of $1.5$ appearing in~\cref{amctuned} for the tuned charm quark mass was chosen based on the results in Table~III of~\cite{Hatton:2020qhk}, and we take the physical $J/\psi$ mass to be $M_{J/\psi}^\mathrm{phys}=3.0969\mathrm{GeV}$~\cite{Workman:2022ynf} where we do not include the negligibly small uncertainty.

We use the simple heuristic forms $M_{\mathrm{pole},i} = M_{\mathrm{pole},i}^\mathrm{phys} + M_{H_c}^\mathrm{latt} - M_{B_c}^\mathrm{phys}$ and $M_H = M_B^\mathrm{phys} + M_{H_c}^\mathrm{latt} - M_{B_c}^\mathrm{phys}$ to approximate the heavy quark mass dependence of the meson masses we do not compute directly on the lattice. These forms ensure the correct values are reached in the physical continuum by setting $M_{H_c}^\mathrm{latt}=M_{B_c}^\mathrm{phys}$. These forms also ensure that 
\begin{align}
M_{\mathrm{pole},i} < t_+ = M_H +M_{D^*}
\end{align}
for all values of $M_{H_c}$. This is only true up to factors behaving as inverse powers of the heavy quark mass. However, since the Blaschke factors approach unity as $M_{\mathrm{pole},i}\to t_+$, the effect of neglecting any poles potentially crossing the pair production threshold is expected to be very small. The values of $M_{\mathrm{pole},i}^\mathrm{phys}$, $M_B^\mathrm{phys}$, and $M_{B_c}^\mathrm{phys}$ that we use are given in~\cref{tab:physmasses}.
We include physical dependence on $m_h$ in the coefficients
\begin{align}
a^{Y}_n = \alpha^{Y}_n \times \Big(1+\sum_{j\neq 0}^3 b_n^{Y,j}&\Delta_{h}^{(j)} \Big)
\end{align}
where $\Delta_{h}^{(j= 0)}=1$ and
\begin{align}\label{deltaMH}
\Delta_{h}^{(j\neq 0)}=\left(\frac{\Lambda_\mathrm{QCD}}{2M_{H}}\right)^j-\left(\frac{\Lambda_\mathrm{QCD}}{2M_{B}^\mathrm{phys}}\right)^j,
\end{align}
where we use $M_H$ as a proxy for $m_h$. Throughout our physical-continuum extrapolation we take $\Lambda_\mathrm{QCD}=0.5\mathrm{GeV}$. We use Gaussian priors of $0(1)$ for $b_n^{Y,j}$ and uniform priors for each $\alpha_n^Y$, corresponding to the the weak unitarity bound $|a_n^Y|\leq 1$ at the physical point $m_h=m_b$. When evaluating the factor of $\overline{m}_h(\overline{m}_h)$ appearing in the susceptibilities we neglect the very small uncertainties of either $\overline{m}_c(3\mathrm{GeV})$ or $\alpha_s$, although we do include this uncertainty when running the tensor renormalisation factors to scale $\overline{m}_h(\overline{m}_h)$. The physical point $m_h=m_b$ is specified by setting $M_{H_c}=M_{B_c}^\mathrm{phys}$ in~\cref{eq:ff}.

The numerical values of the physical masses that we use in our parameterisation are listed in~\cref{tab:physmasses}. Note that we do not include uncertainties for these masses.

\begin{table}
\centering
\caption{Physical masses of mesons, $X$, used in our physical continuum fit function, as well as pole masses, $M_{\mathrm{pole},i}^\mathrm{phys}$, for each form factor, $Y$. \label{tab:physmasses}}
\begin{tabular}{ c | c }
\hline
$X$ & $M_X^\mathrm{phys}/$GeV \\
\hline
${J/\psi}$ & 3.0969\\
${B_c}$    & 6.2745\\
${D^*}$    & 2.010\\
${B}$      & 5.280\\
${\pi}$    & 0.13957\\
\hline
$Y$ & $M_{\mathrm{pole},i}^\mathrm{phys}/$GeV \\
\hline
$A_{0}$ & 6.2749, 6.872, 7.25\\
$A_{1,12}$, $T_{2,23}$ & 6.745, 6.75, 7.145, 7.15\\
$V$, $T_1$ & 6.335, 6.926, 7.02, 7.28\\
\hline
\end{tabular}
\end{table}

\subsection{Kinematical Constraints}

The form factors must satisfy kinematical constraints at $q^2=0$,
\begin{align}\label{kinz}
A_0(q^2=0) =& \frac{M_{H_c}+M_{J/\psi}}{2M_{J/\psi}}A_1(q^2=0)\nonumber\\
&-\frac{M_{H_c}-M_{J/\psi}}{2M_{J/\psi}}A_2(q^2=0)\nonumber\\
T_1(q^2=0) =& T_2(q^2=0)\nonumber\\
\end{align}
and at $q^2_\mathrm{max}=(M_{H_c}-M_{J/\psi})^2$,
\begin{align}\label{kinmax}
&A_{12}(q^2=q^2_\mathrm{max}) = \nonumber\\
&\frac{\left(M_{H_c}+M_{J/\psi}\right)\left(M_{H_c}^2-M_{J/\psi}^2-q^2_\mathrm{max}\right)}{16M_{H_c}M_{J/\psi}^2}A_1(q^2=q^2_\mathrm{max})\nonumber\\
&T_{23}(q^2=q^2_\mathrm{max}) = \nonumber\\
&\frac{\left(M_{H_c}+M_{J/\psi}\right)\left(M_{H_c}^2+3M_{J/\psi}^2-q^2_\mathrm{max}\right)}{ 8M_{H_c}M_{J/\psi}^2}T_2(q^2=q^2_\mathrm{max}).
\end{align} 

The kinematical constraints in~\cref{kinz,kinmax} are valid in the continuum limit for all of the unphysical values of $M_{H_c}$ in our lattice calculation. For $A_{12}$ and $A_0$ we impose the constraints by fixing the first physical coefficients in \cref{eq:ff} to 
\begin{align}
a_0^{A_{12}} =& \Big[P^{A_{12}}(q^2_\mathrm{max})\phi^{A_{12}}(q^2_\mathrm{max})\nonumber\\
&\times A_1(q^2_\mathrm{max})\frac{(M_{H_c}+M_{J/\psi})(M_{H_c}^2-M_{J/\psi}^2-q^2_\mathrm{max})}{16M_{H_c}M_{J/\psi}^2} \nonumber\\
&- \sum_{n=1}^4 a_n^{A_{12}} p_n(z(q^2_\mathrm{max}))\Big]\frac{1}{p_0(z(q^2_\mathrm{max}))}, \nonumber\\
a_0^{A_{0}} =& \Big[P^{A_{0}}(0)\phi^{A_{0}}(0) \nonumber\\
&\times\Big(A_1(0)\frac{M_{H_c}+M_{J/\psi}}{M_{J/\psi}}-A_2(0)\frac{M_{H_c}-M_{J/\psi}}{M_{J/\psi}} \Big)\nonumber\\
&- \sum_{n=1}^4 a_n^{A_{0}} p_n(z(0))\Big]\frac{1}{p_0(z(0))},
\end{align}
respectively, where $A_1(t_-)$, $A_1(0)$, and $A_2(0)$ are evaluated using \cref{eq:ff} without the mistuning terms $\mathcal{N}_n$, i.e. with $\mathcal{N}_n=1$. Note that this implementation enforces the constraints in the continuum limit exactly for all values of $m_h$. We impose the constraints for $T_1$ and $T_{23}$ similarly by fixing the coefficients
\begin{align}
a_0^{T_{23}} =& \Big[P^{T_{23}}(q^2_\mathrm{max})\phi^{T_{23}}(q^2_\mathrm{max})\nonumber\\
&\times T_2(q^2_\mathrm{max})\frac{(M_{H_c}+M_{J/\psi})(M_{H_c}^2+3M_{J/\psi}^2-q^2_\mathrm{max})}{8M_{H_c}M_{J/\psi}^2} \nonumber\\
&- \sum_{n=1}^4 a_n^{T_{23}} p_n(z(q^2_\mathrm{max}))\Big]\frac{1}{p_0(z(q^2_\mathrm{max}))}, \nonumber\\
a_0^{T_{1}} =& \Big[P^{T_{1}}(0)\phi^{T_{1}}(0) T_2(0) - \sum_{n=1}^4 a_n^{T_{1}} p_n(z(0))\Big]\frac{1}{p_0(z(0))}.
\end{align}

\subsection{Discretisation Effects}
Following~\cite{Harrison:2023dzh}, we fit the matrix elements that we extract on the lattice directly, allowing for discretisation effects using an additive term capturing dependence on momentum and heavy quark mass:

\begin{align}
J^{00}_{nn(\nu,\Gamma)} = J_\mathrm{phys}^{\nu,\Gamma}& \nonumber\\
+ \sum_{j=0}^3\sum_{k+l\neq 0}^2& c^{(\nu,\Gamma),jkl}\Delta_{h}^{(j)} \left({ak}\right)^{2k}\left(\frac{am_h^\mathrm{val}}{\pi}\right)^{2l}.
\end{align}
$J_\mathrm{phys}^{\nu,\Gamma}$ is computed from the form factors determined using~\cref{eq:ff} combined with~\cref{formfactors} and~\cref{relnorm,formfactorsQCDSM,formfactorsQCDNP,helicityffsA12T23}. Note that unlike our previous calculations, we omit powers of $(a\Lambda_\mathrm{QCD}/\pi)^2$, which are generated by the mixed terms such as $(am_h^\mathrm{val}/\pi)^2(\Lambda_\mathrm{QCD}/M_{H})^2\approx (a\Lambda_\mathrm{QCD}/\pi)^2$. We truncate the sum over discretisation effects above order $a^4$, since, even for our largest masses $am_h=0.8$ and largest values of $ak$, $(am_h/\pi)^6$ and $(ak)^6$ are both very small. We investigate the impact of these choices in~\cref{contstabsec}, where we see that varying the maximum order of $a^2$, $\Lambda_\mathrm{QCD}/M_H$ or $p_n(z)$ terms included in our physical continuum fit function has a negligible impact on the resulting form factors. We take Gaussian priors of $0(1)$ for each $c^{(\nu,\Gamma),jkl}$.

\subsection{Continuum Results}
Our continuum form factor results for $B_c\to J/\psi$ are plotted in~\cref{ffcorrectedplots}, together with our lattice data, both at the physical bottom quark mass, $m_h=m_b$, as well as for the unphysically light values $m_h =4.12m_c,~2.91m_c$, and $1.78m_c$ corresponding to the largest heavy quark masses used on sets 3, 2 and 1 respectively. Note that to determine the (purely illustrative) values of $M_{H_c}$ corresponding to these unphysical heavy quark masses, we fit our lattice results for the $H_c$ masses using a heuristically chosen function of $\overline{m}_h(\overline{m}_h)$, including $am_c$ and $am_h$ discretisation effects, as well as scale dependence through $\alpha(\overline{m}_h)$. The fit function we use is
\begin{align}
M_{H_c}^\mathrm{latt} = &\overline{m}_h\mathcal{N}'_{\overline{m}_h}+A\mathcal{N}'_A + \sum_{n=1}^6 \left(\frac{\Lambda_\mathrm{QCD}}{\overline{m}_h}\right)^{n/2}\mathcal{N}'_n\nonumber\\
+&b_0 (am_c/\pi)^2+b_1 (am_c/\pi)^4\nonumber\\
+&c_0 (am_h/\pi)^2+c_1 (am_h/\pi)^4
\end{align}
where each $\mathcal{N}'_X$ is defined analogously to~\cref{mistuning}, with the addition of a scale dependent term $E_X(\alpha_s( \overline{m}_h)/\pi)$. For each $A$, $E_X$, $b_n$, $c_n$, sea quark and valence charm quark mistuning coefficients, we take Gaussian priors of $0(4)$. Note that this fit is used purely to illustrate the heavy quark mass dependence of the form factors extracted from our fit to~\cref{eq:ff}, and does not enter the subsequent analysis.
In order to illustrate the consistency between our data and fit function, in~\cref{ffcorrectedplots} we plot our data corrected using the dependence on $(ak)^2$, $(am_h)^2$ and sea and valence quark mistunings determined from our fit so that the data points we plot are 
\begin{align}\label{correction}
F^\mathrm{corrected}&(a,am_h,M_{H_c})=F^\mathrm{data}(a,am_h,M_{H_c}) \nonumber\\
-\Big[F^\mathrm{fit}(a^2,&am_h,\delta,M_{H_c})-F^\mathrm{fit}(0,0,0,M_{H_c})\Big].
\end{align}
Note that this requires converting the matrix elements for each ensemble, computed using our fit function and posteriors, back to form factors. We see that our data agree well with our chiral continuum fit function. 

The dispersive parameterisation,~\cref{eq:ff}, is based on the fact that $\int_{\mathcal{C}_\alpha}|P^Y(z,t_+,t_0)\phi^Y(z,t_+,t_0)F^Y|^2\leq 1$, where the integral is over the arc of the unit circle defined in~\cref{orthopols}, and hence the combination $P^Y(z,t_+,t_0)\phi^Y(z,t_+,t_0)F^Y$ may be expressed as a sum over orthonormal polynomials in $z$~\cite{Gubernari:2023puw}. In~\cref{ffPphiplots} we plot our data and continuum form factors multiplied by $P^Y(z,t_+,t_0)\phi^Y(z,t_+,t_0)$, to illustrate the simple dependence of $P^Y(z,t_+,t_0)\phi^Y(z,t_+,t_0)F^Y$ on $z$. We see that in this space, our data has very simple dependence on $z$, as expected.

\subsection{Error Budget}

Here we give a breakdown of the total uncertainties for our physical continuum form factors from different sources, across the full $q^2$ range. For each $q^2$, we compute the partial variance with respect to different priors, corresponding to systematic uncertainties remaining from different elements of our fit function, as well as with respect to the statistical uncertainty of data on each ensemble. We plot the partial variances, normalised by the total variance, in~\cref{fferrorbandplots}. There, we see that statistical uncertainties are the largest source of uncertainty, mostly coming from set 4. Note that since we do not include explicit $(a\Lambda_\mathrm{QCD}/\pi)^2$ terms, which are instead generated by mixed $(\Lambda_\mathrm{QCD}/M_H)^2$ and $(am_h/\pi)^2$ terms, these $(a\Lambda_\mathrm{QCD}/\pi)^2$ uncertainties are included in~\cref{fferrorbandplots} in the yellow `mixed' band.

Compared to the results of~\cite{Harrison:2020gvo}, where $(\Lambda_\mathrm{QCD}/M_H)^2$ and $(am_h/\pi)^2$ contributed the dominant systematic uncertainties, we see a substantial reduction in the relative size of these uncertainties compared to other effects. In terms of statistical uncertainties, we also see a significant reduction in the impact of set 3, which previously contributed the dominant statistical uncertainty. These changes result from the addition of set 4 in this work, which now anchors the physical continuum fit at $m_h=m_b$ and across the kinematical range. With the reduction in total uncertainty of our form factors, we also see increased relative importance of chiral and mass mistuning effects compared to~\cite{Harrison:2020gvo}, owing to the reduced overall uncertainty.

\section{Discussion}
\label{sec:disc}
\begin{figure}
\includegraphics[scale=0.3]{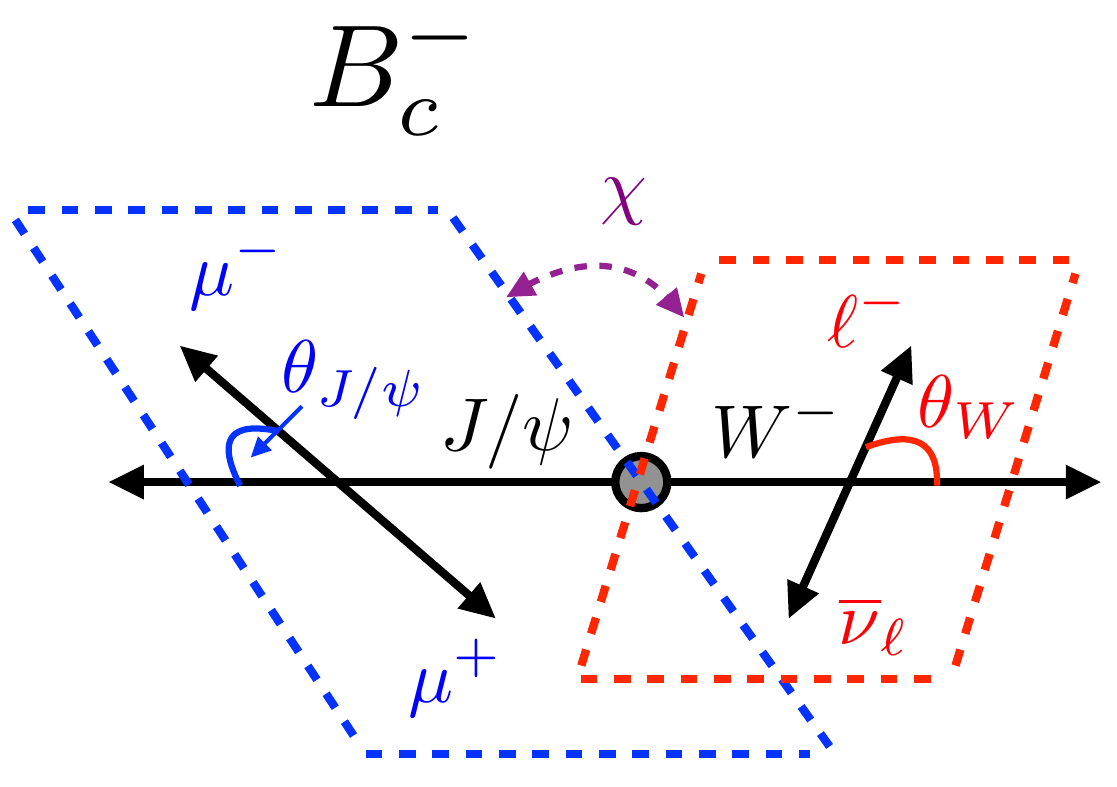}
\caption{\label{angles}Conventions for the angular variables entering the differential decay rate. The conventions used here are the same as in~\cite{Harrison:2020gvo}.}
\end{figure}

\begin{figure}
\centering                            
\includegraphics[width=0.5\textwidth]{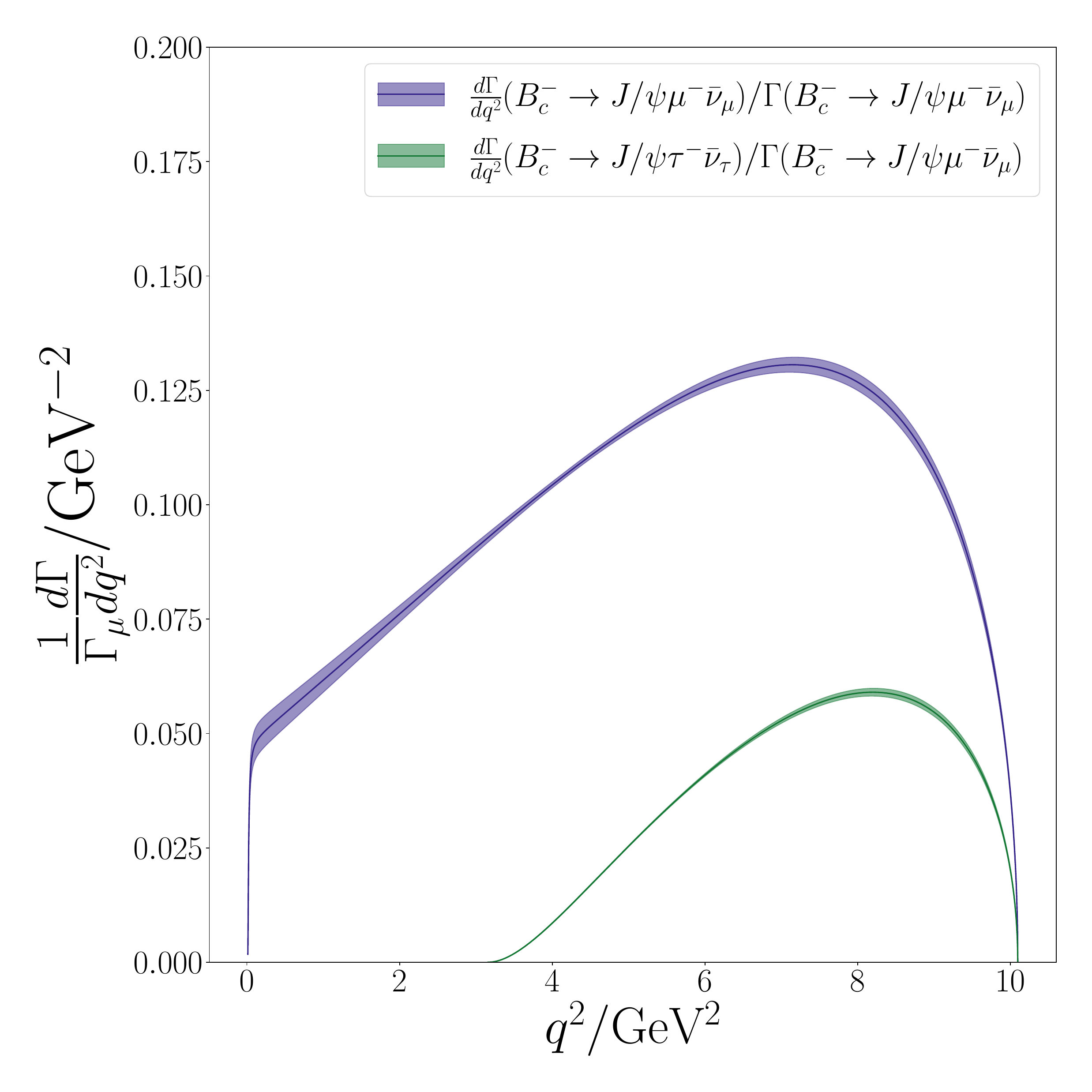}
\caption{\label{dgammadqsq} Predicted differential decay rate, normalised by the total decay rate for the semimuonic mode. The blue curve shows $B_c^-\to J/\psi \mu^-\bar{\nu}$, and the green curve shows the semitauonic mode, $B_c^-\to J/\psi \tau^-\bar{\nu}$.}
\end{figure}

\subsection{Standard Model Phenomenology}
\label{subsec:pheno}

The full differential rate for $B_c^-\to J/\psi(\to \mu^+\mu^-)\ell^-\bar{\nu}$ is given by~\cite{Cohen:2018vhw}
\begin{align}
\label{eq:diffrate}
\frac{d^4\Gamma(B_c^-\rightarrow J/\psi(\rightarrow\mu^+\mu^-)\ell^-\overline{\nu})}{d\cos(\theta_{J/\psi})d\cos(\theta_{W})d\chi dq^2}&=\nonumber\\
 \mathcal{B}(J/\psi\rightarrow\mu^+\mu^-)\mathcal{N}\sum_ik_i&(\theta_W,\theta_{J/\psi},\chi)\mathcal{H}_i(q^2) 
\end{align}
where $\theta_W$ is the angle between the lepton and $W^-$ momenta in the $W^-$ rest frame, $\theta_{J/\psi}$ is the angle between the $\mu^-$ and $J/\psi$ momenta, and $\chi$ is the angle between the $\mu^+\mu^-$ and $\ell\bar{\nu}$ planes. These angles are shown in~\cref{angles}. $k_i(\theta_W,\theta_{J/\psi},\chi)$ are known kinematic factors, $\mathcal{H}_i(q^2)$ are helicity amplitudes, and $\mathcal{N}$ is a normalisation factor
\begin{equation}
\mathcal{N}=\frac{G_F^2}{(4\pi)^4}|\eta_{EW}|^2|V_{cb}|^2\frac{ 3(q^2-m_\ell^2)^2|\vec{p^\prime}| }{8M_{B_c}^2 q^2}
\end{equation}
where $\eta_\mathrm{EW}$ is a structure-independent electroweak 
correction~\cite{Sirlin:1981ie}. 
The $k_i$ and $\mathcal{H}_i$ are given in Table~1 of~\cite{Harrison:2020nrv} for the SM case. 
The SM differential rate with respect to $q^2$ is given by
\begin{align}
&\frac{d\Gamma}{dq^2} = \mathcal{N}\times\frac{64\pi}{9}\Big[\left({H_-}^2+{H_0}^2+{H_+}^2\right) \nonumber\\
+&\frac{{m_\ell^2}}{2q^2}{ \left({H_-}^2+{H_0}^2+{H_+}^2+3 {H_t}^2\right)}\Big],
\end{align}
where the helicity amplitudes are defined in terms of the form factors in the QCD basis as~\cite{RevModPhys.67.893}
\begin{align}
H_\pm(q^2) =& (M_{B_c}+M_{J/\psi})A_1(q^2) \mp \frac{2M_{B_c}|\vec{p'}|}{M_{B_c}+M_{J/\psi}}V(q^2),\nonumber\\
H_0(q^2) =& \frac{1}{2M_{J/\psi} \sqrt{q^2}} \Big(-4\frac{M_{B_c}^2{|\vec{p'}|}^2}{M_{B_c}+M_{J/\psi}}A_2(q^2)\nonumber\\
&  +  (M_{B_c}+M_{J/\psi})(M_{B_c}^2 - M_{J/\psi}^2 - q^2)A_1(q^2) \Big),\nonumber\\
H_t(q^2) =& \frac{2M_{B_c}|\vec{p'}|}{\sqrt{q^2}}A_0(q^2).\label{helicityamplitudes}
\end{align}
From these expressions, it is straightforward to compute the differential decay rates for the semimuonic and semitauonic cases, shown in~\cref{dgammadqsq} normalised by the total semimuonic rate. We also calculate the total decay rates, for which we find
\begin{align}
\frac{\Gamma(B_c^-\to J/\psi\tau^-\bar{\nu})}{|\eta_\mathrm{EW}V_{cb}|^2}=&{\gammatau},\nonumber\\
\frac{\Gamma(B_c^-\to J/\psi\mu^-\bar{\nu})}{|\eta_\mathrm{EW}V_{cb}|^2}=&{\gammamu},\nonumber\\
\frac{\Gamma(B_c^-\to J/\psi e^-\bar{\nu})}{|\eta_\mathrm{EW}V_{cb}|^2}=&{\gammae}.
\end{align}
These values are in excellent agreement with the values previously computed in~\cite{Harrison:2020gvo,Harrison:2020nrv}, and the uncertainties are roughly a factor of 2 smaller. 

The structure-independent electroweak correction is given by $\eta_\mathrm{EW}=1+\alpha_s\mathrm{log}(M_Z/\mu)/\pi$~\cite{Sirlin:1981ie}. Evaluating this at the scale $\mu=M_{B_c}$, and estimating the uncertainty by varying this scale by a factor of 2 gives $\eta_\mathrm{EW}=1.0062(16)$. Using this value, together with the global fit result of~\cite{Workman:2022ynf}, $|V_{cb}|=41.82^{+0.85}_{-0.74}$, gives
\begin{align}
\Gamma(B_c^-\to J/\psi\mu^-\bar{\nu})=&{\gammamufull}\nonumber\\
                                     =&{\gammamufullGeV},
\end{align}

where the first uncertainty is from our lattice calculation and the second is from $|\eta_\mathrm{EW}V_{cb}|^2$. We can combine this value with the experimental average value for the $B_c$ lifetime, $\tau_{B_c}=0.510(9)\times 10^{-12}\mathrm{s}$~\cite{Workman:2022ynf}, to give
\begin{align}
\mathrm{Br}(B_c^-\to J/\psi\mu^-\bar{\nu})=&{\Brmufull}.
\end{align}
Here the first uncertainty is from our lattice calculation, the second from $|\eta_\mathrm{EW}V_{cb}|^2$ and the third from $\tau_{B_c}$. The Particle Data Tables~\cite{Workman:2022ynf} give the $B_c^-\to J/\psi\mu^-\bar{\nu}$ branching fraction multiplied by the factor $\mathrm{B}(\bar{b}\to B_c^+)$, corresponding to the probability that a $b$ quark hadronises to a $B_c$ state, including possible excitations. We can therefore use the value of $\mathrm{B}(\bar{b}\to B_c^+)\times \mathrm{Br}(B_c^-\to J/\psi\mu^-\bar{\nu})=8.8(1.0)\times 10^{-5}$ from~\cite{Aaltonen:2016dra}, given in~\cite{Workman:2022ynf}, together with our value of $\mathrm{Br}(B_c^-\to J/\psi\mu^-\bar{\nu})$ to determine $\mathrm{B}(\bar{b}\to B_c^+)$. We find

\begin{align}
\mathrm{B}(\bar{b}\to B_c^+)=&{\btoBc}
\end{align}
Where the uncertainties are from our lattice calculation, $|\eta_\mathrm{EW}V_{cb}|^2$, $\tau_{B_c}$ and $\mathrm{B}(\bar{b}\to B_c^+)\times \mathrm{Br}(B_c^-\to J/\psi\mu^-\bar{\nu})$ respectively.

We also determine the ratio of the semimuonic and semielectronic modes $\Gamma_\mu/\Gamma_e={\RATIOEMU}$, which we find to be in good agreement with our previous calculation. Our new result for the important lepton flavour universality ratio
\begin{align}
R(J/\psi)=\frac{\Gamma(B_c^-\to J/\psi\tau^-\bar{\nu})}{\Gamma(B_c^-\to J/\psi\mu^-\bar{\nu})}={\RJpsi}
\end{align}
is in excellent agreement with the value computed in~\cite{Harrison:2020nrv}, $R^\mathrm{2020}(J/\psi)=0.2582(38)$, but note that it is substantially more precise.

We also compute the lepton polarisation asymmetry, $\mathcal{A}_{\lambda_\ell}$, the longitudinal polsarisation fraction, $F_{L}^{J/\psi}$, and the forward-backward asymmetry, $\mathcal{A}_{FB}$. These are defined as~\cite{Becirevic:2019tpx}
\begin{align}
\label{angasymeq}
\mathcal{A}_{\lambda_\ell}(q^2) =& \frac{d\Gamma^{\lambda_\ell=-1/2}/dq^2-d\Gamma^{\lambda_\ell=+1/2}/dq^2}{d\Gamma/dq^2},\nonumber\\
F_{L}^{J/\psi}(q^2) =& \frac{d\Gamma^{\lambda_{J/\psi}=0}/dq^2}{d\Gamma/dq^2},\nonumber\\
\mathcal{A}_\mathrm{FB}(q^2) =& -\frac{1}{d\Gamma/dq^2} \frac{2}{\pi}\int_0^\pi\frac{d\Gamma}{dq^2d\cos(\theta_W)}\cos(\theta_W)d\theta_W.
\end{align}
Using the SM form factors computed in this work, we find
\begin{align}
A_{\lambda_\tau} =         &   \almtau,                     \nonumber\\
F_L^{J/\psi}     =         &   \Fljpsi,                     \nonumber\\
\mathcal{A}_\mathrm{FB} = &   \Afb.
\end{align}
Both $\mathcal{A}_\mathrm{FB}$ and $F_L^{J/\psi}$ are in excellent agreement with those computed in~\cite{Harrison:2020nrv}, however $A_{\lambda_\tau}$ is roughly $1\sigma$ lower.

\subsection{Comparison of form factors to previous calculation}

\begin{figure}
\centering                            
\includegraphics[width=0.5\textwidth]{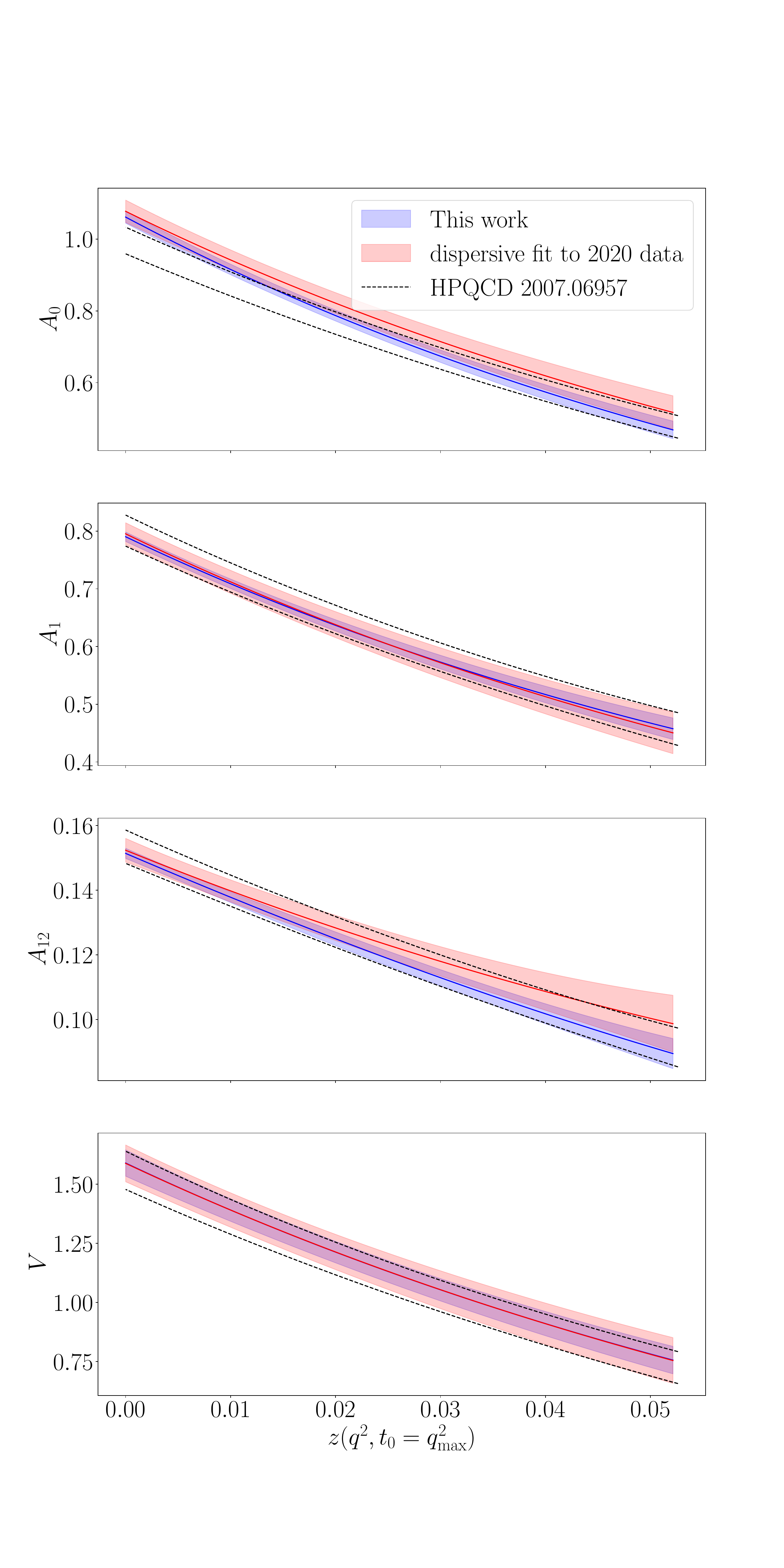}
\caption{\label{ffcomparisonplots} Comparison of the results of this work (blue line and error band) to the result of fitting the data used in~\cite{Harrison:2020gvo} using the fit function of this work (red line and error band). The black dashed lines correspond to the $\pm 1\sigma$ confidence interval of our previous result~\cite{Harrison:2020gvo}.}
\end{figure}

We compare the results of this work for the SM form factors $V$, $A_0$, $A_1$, and $A_{12}$ to those of~\cite{Harrison:2020gvo}. In that work, the physical continuum extrapolation used a simple expansion in $z$, combined with the Blaschke factors $\prod_{i}z(q^2,t_+,M_{\mathrm{pole},i}^2)$, but without the inclusion of outer functions and with heuristically chosen priors of $0\pm 1$ for the coefficients of the expansion. In~\cref{ffcomparisonplots} we show the results of this work together with the results of~\cite{Harrison:2020gvo}, as well as the results obtained applying the fit function of~\cref{eq:ff} to the lattice data from~\cite{Harrison:2020gvo}. We see that the fit function used in this work results in a shift of $\approx 1\sigma$ in $A_0$ close to zero recoil, but very consistent results are found for the remaining form factors. We also see that the results of this work are compatible with the previous work for all form factors across the kinematic range, with the exception of a $1\sigma$ shift in $A_0$ close to zero recoil.

\subsection{Comparison to NRQCD tensor-SM form factor relations}
\label{sec:NRQCD}

\begin{figure}
\centering                            
\includegraphics[width=0.5\textwidth]{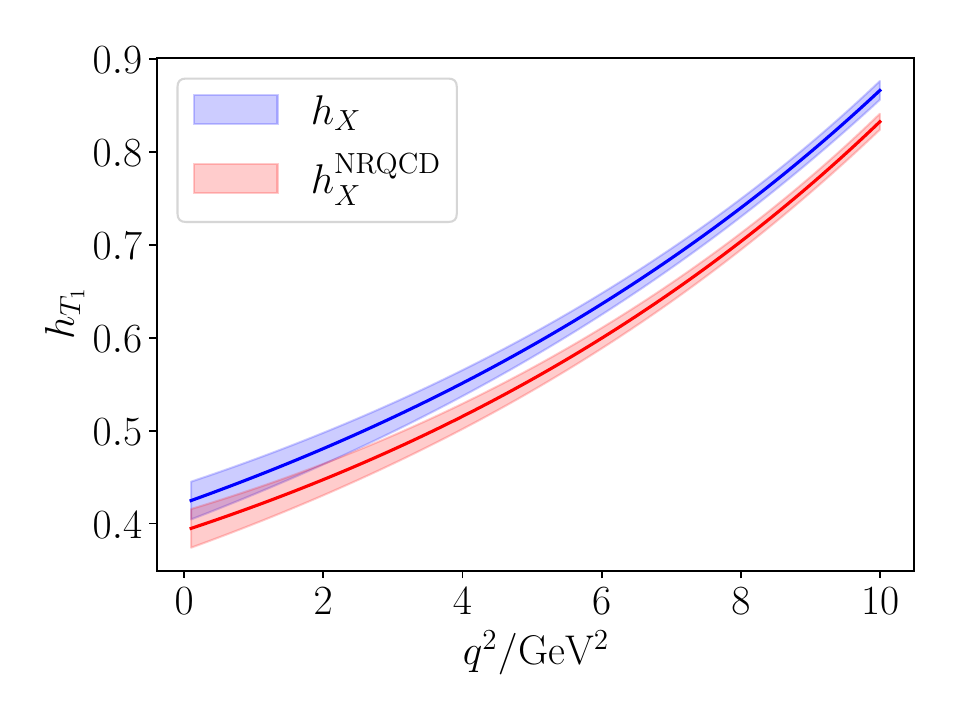}\\
\includegraphics[width=0.5\textwidth]{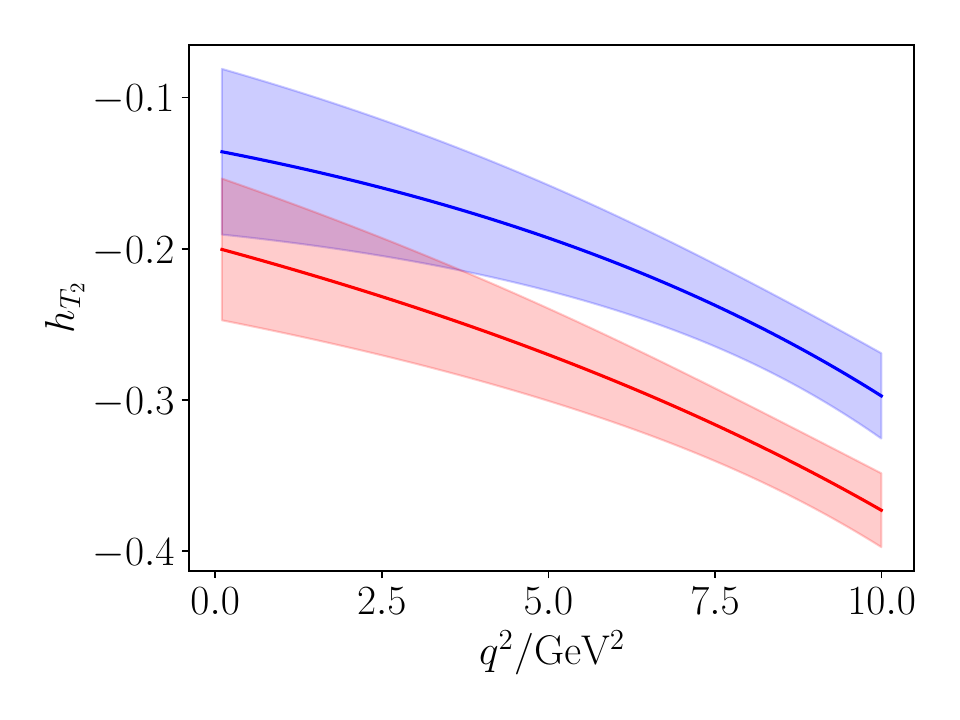}\\
\includegraphics[width=0.5\textwidth]{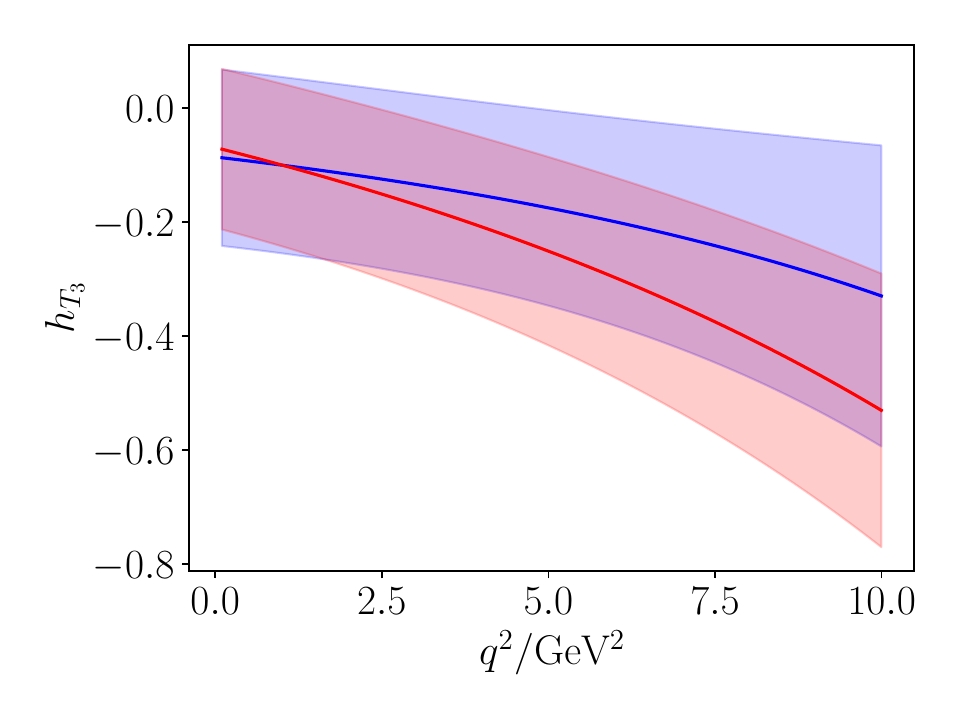}
\caption{\label{NRQCDrelations} Here we show the tensor form factors resulting from our lattice QCD calculation (blue) together with the results derived using the $\mathcal{O}(1/m)$ nonrelativistic QCD results of~\cite{Colangelo:2022lpy}, given in~\cref{NRQCDeq}, together with the SM form factors of this work (red). We see substantial violations of the relations.}
\end{figure}

Here we use our SM and tensor form factors to test the $\mathcal{O}(1/m)$ nonrelativistic QCD results of~\cite{Colangelo:2022lpy}, which gives expressions for the tensor form factors in terms of the SM form factors. These relations, which are expressed in terms of the form factors in the HQET basis defined in~\cref{formfactors}, are
\begin{align}
h_{T_1}^\mathrm{NRQCD}(w)&=\frac{1}{2} \Big((1+w)h_{A_1}(w)-(w-1)h_V(w) \Big), \label{NRQCDeq}\\
h_{T_2}^\mathrm{NRQCD}(w)&=\frac{1+w}{2 (m_b + 3 m_c)}\Big((m_b - 3 m_c) h_{A_1}(w),  \nonumber\\
&~~~+ 2 m_c (h_{A_2}(w)+ h_{A_3}(w)), \nonumber\\
&~~~- (m_b - m_c) h_V(w)\Big) \nonumber\\
h_{T_3}^\mathrm{NRQCD}(w)&= h_{A_3}(w)  -  h_V(w). \nonumber
\end{align}
We use the ratio of charm and bottom quark pole masses $u^\mathrm{pole}=m_c^\mathrm{pole}/m_b^\mathrm{pole}=0.33$ to evaluate these expressions, together with our lattice QCD results for the SM form factors. The resulting tensor form factors are compared to our pure lattice QCD results for the tensor form factors in~\cref{NRQCDrelations}. We see that the NRQCD relations of~\cref{NRQCDeq} are significantly violated. This indicates that the higher order $\mathcal{O}(1/m^2)$ terms in the NRQCD expansion are relevant at this level of precision. This mirrors the situation for the related decay $B\to D^*$, where $\mathcal{O}(1/m_c^2)$ terms in the heavy quark expansion are required to satisfactorily describe lattice results~\cite{Bordone:2019vic}.

\subsection{Fits using only high-$q^2$ data}

\begin{figure*}
\centering                            
\includegraphics[width=1\textwidth]{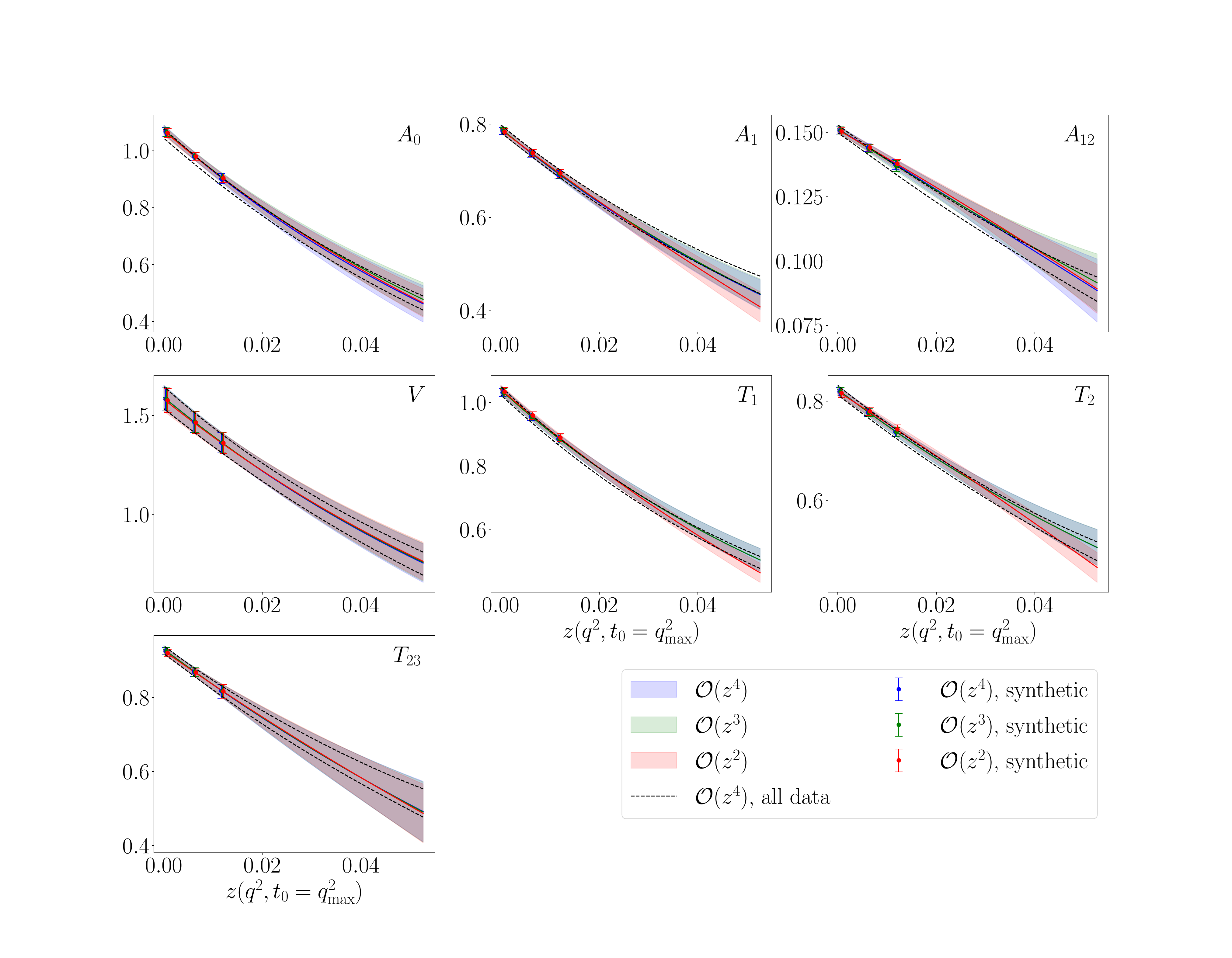}
\caption{\label{ffcomparisonplotsTRUNC} Here we show the synthetic data points generated from fits to our lattice data from just the $z<0.015$ region for fits to~\cref{eq:ff} truncated at orders $z^2$, $z^3$ and $z^4$ in red, green and blue respectively. We also show the results of fitting these synthetic data points using~\cref{eq:ff} to extrapolate the form factors into the low-$q^2$ region. We show the $1\sigma$ uncertainty interval of our main results using our full data set as the black dashed lines.}
\end{figure*}

\begin{table}
\centering
\caption{\label{A1synthpoints} Synthetic data points for the $A_1$ form factor resulting from fits to data in the high-$q^2$ region, $z<0.015$, with different maximum powers of~$z$ included in~\cref{eq:ff}.}
\begin{tabular}{ c c c c | c }
\hline
$A_1(q^2/\mathrm{GeV}^2)$ & $A_1(10)$ & $A_1(9)$ & $A_1(8)$  & $A_1(0)$ \\\hline
$\mathcal{O}(z^2)$&0.7836(73)&0.7381(72)&0.6946(87)&0.408(33)\\
$\mathcal{O}(z^3)$&0.7857(75)&0.7375(74)&0.6931(89)&0.436(33)\\
$\mathcal{O}(z^4)$&0.7855(75)&0.7366(75)&0.6916(90)&0.435(32)\\
\hline
\end{tabular}
\end{table}

In this section, we investigate the results of fitting just our high-$q^2$ data, while varying the truncation order, $n$, in our fit function,~\cref{eq:ff}. As is typical for calculations with lattice data only in a limited $q^2$ range, we produce synthetic data points in the high-$q^2$ region which we fit with the dispersive parameterisation  to extrapolate the form factors to the low-$q^2$ region~\cite{FermilabLattice:2021cdg,Aoki:2023qpa}. We first restrict the input data to $z<0.015$, corresponding to roughly the first $1/3$ of the full kinematical range closest to zero recoil, which is typical of recent lattice QCD calculations of the $B\to D^*$ form factors~\cite{FermilabLattice:2021cdg,Aoki:2023qpa} that do not use the HISQ action for valence quarks. We then perform physical continuum extrapolations using our full fit function,~\cref{eq:ff}, including up to $z^4$ terms, as well as fits including only up to $z^3$ and $z^2$ terms. For each of these fits we generate synthetic data points in the $z<0.015$ region which we then fit using our full physical continuum dispersive parameterisation, in order to compare the extrapolated behaviour in the low-$q^2$ region. 

We plot the results for each truncation order in~\cref{ffcomparisonplotsTRUNC}. There, we see that the $\mathcal{O}(z^4)$ fit to data in just the high-$q^2$ region agrees well with our full results when extrapolated using the dispersive parameterisation. We also see that for $A_0$, $A_{12}$, $V$ and $T_{23}$ the form factors are consistent across the full range. However, for $A_1$, $T_1$ and $T_2$ the synthetic data points, which are fully consistent, result in significant differences when extrapolated to the low-$q^2$ region. Note that all the fits to the reduced range of data considered in this subsection have $\chi^2/\mathrm{dof}< 1$.

For phenomenology, this effect is particularly relevant for the form factor $A_1$~(see~\cref{subsec:pheno}). We give the values of the synthetic data points for $A_1$ for fits with different truncation orders in~\cref{A1synthpoints}, together with the extrapolated values at $q^2=0$. We see there that when going from $\mathcal{O}(z^3)$ to $\mathcal{O}(z^2)$ there is only a very slight change in the synthetic data points, but the extrapolated value at $q^2=0$ changes by almost $1\sigma$, resulting in tension with the value obtained from the full fit. This observation highlights the importance of including all systematic uncertainties associated with kinematics when performing extrapolations from high-$q^2$ to low-$q^2$. 

In recent lattice QCD form factor calculations for $B\to D^*$, the chiral continuum extrapolation is typically repeated including additional kinematic and discretisation terms, in order to verify that continuum results do not change significantly and that all relevant systematic uncertainties have been included~\cite{FermilabLattice:2021cdg,Harrison:2023dzh,Aoki:2023qpa}. Our findings here suggest that lattice form factor calculations working in a limited kinematical range should also verify that the low-$q^2$ form factors (determined by fitting synthetic data with a dispersively-bounded parameterisation) are insensitive to the inclusion of higher order terms in $z$ in the chiral continuum fit function.

\section{Conclusions and outlook}
\label{sec:conc}
We have computed the full set of Standard Model and tensor form factors for $B_c\to J/\psi$ semileptonic decay, using the heavy-HISQ approach in which data are generated using a variety of heavy quark masses and extrapolated to the physical bottom quark mass. 
The calculation presented here improves significantly on our earlier work~\cite{Harrison:2020gvo}, including additional lattice data for the $B_c\to J/\psi$ form factors on two new ensembles, one with physically light up and down quarks and $a\approx 0.06 \mathrm{fm}$ and one with $a\approx 0.03\mathrm{fm}$ on which we are able to reach the physical bottom quark mass. Our calculation achieves excellent coverage of the kinematical range, with data on the $a\approx 0.03\mathrm{fm}$ and $a\approx 0.045\mathrm{fm}$ ensembles spanning the full $q^2$-range. Utilising our recent lattice QCD calculation of the $\bar{h}c$ susceptibilities as a function of $u=m_c/m_h$, we use the full dispersive parameterisation including outer functions and Blaschke factors to extrapolate our data to the physical bottom quark mass in the chiral-continuum limit. Our results are found to be consistent with our previous calculation, and roughly a factor of 2 more precise. 

We see only minor differences between the form factors determined in this work and in the previous work, amounting to at most $1\sigma$ for $A_0$ in the high-$q^2$ (zero recoil) region, and we have verified that applying the dispersive parameterisation of this work to the lattice data used in our previous work results in form factors that are compatible with this work for all $q^2$ values. We have included a comprehensive breakdown of the uncertainties entering our form factors, where we see a reduction in systematic uncertainties associated with the heavy quark mass dependence and heavy quark discretisation effects relative to~\cite{Harrison:2020gvo}.

We have used our form factor results to determine updated values for important Standard Model observables, such as the differential decay rates and the total decay rates for both the semimuonic and semitauonic modes. The updated value of $R(J/\psi)=\RJpsi$ that we find here is in good agreement with the previous result, and substantially more precise. We have provided updated values of $F_L^{J/\psi}     =   \Fljpsi$, and $\mathcal{A}_\mathrm{FB} =   \Afb$ which also show good agreement with our previous numbers. The value of $A_{\lambda_\tau} = \almtau$ that we find differs from our previous result by $\approx 1\sigma$, owing to the shift in the $A_0$ form factor close to zero recoil.

We have also presented the first lattice QCD results for the $B_c\to J/\psi$ tensor form factors. Combined with precise experimental data for $B_c\to J/\psi\ell\nu$ expected in the future, these will enable the determination of constraints on new physics complementary to those determined from $B\to D^{(*)}$ and $B_s\to D_s^{(*)}$ decays. We have also used our results for the $B_c\to J/\psi$ tensor form factors to test the $\mathcal{O}(1/m)$ nonrelativistic QCD results of~\cite{Colangelo:2022lpy}, relating the tensor form factors to the SM form factors. We find significant violations of these relations, particularly near zero recoil, suggesting that $\mathcal{O}(1/m_c^2)$ effects are important at this level of precision.

To test the extrapolation of high-$q^2$ lattice data into the low-$q^2$ region, we have performed physical continuum fits to lattice data restricted to the high-$q^2$ region, including terms up to and including $\mathcal{O}(z^2)$, $\mathcal{O}(z^3)$, and $\mathcal{O}(z^4)$. As expected, we found that the high-$q^2$ fits were all compatible with one another, and with our main result, in the high-$q^2$ region. We used these fits to produce synthetic data points in the high-$q^2$ region, which were then extrapolated to the low-$q^2$ region using our continuum dispersive parameterisation. The $\mathcal{O}(z^4)$ and $\mathcal{O}(z^3)$ fits gave compatible values to our main result, while the fit including terms up to only $\mathcal{O}(z^2)$ resulted in $\approx 1\sigma$ differences, despite being compatible in the high-$q^2$ region. These observations suggest that the impact of uncertainties from higher order kinematical terms on the extrapolation to the low-$q^2$ region should be investigated in lattice calculations of related $b\to c$ form factors such as for $B\to D^*$.

\subsection*{\bf{Acknowledgements}}

I am grateful to the MILC Collaboration for the use
of their configurations and code. We thank C. T.H. Davies, C. Bouchard, D. van Dyk, M. Reboud, M. Jung and M. Bordone for useful discussions. Computing was done on
the Cambridge service for Data Driven Discovery (CSD3),
part of which is operated by the University of Cambridge
Research Computing on behalf of the DIRAC HPC Facility
of the Science and Technology Facilities Council (STFC).
The DIRAC component of CSD3 was funded by BEIS
capital funding via STFC capital Grants No. ST/P002307/1
and No. ST/R002452/1 and by STFC operations Grant
No. ST/R00689X/1. DiRAC is part of the national
e-infrastructure. We are grateful to the CSD3 support staff
for assistance. Funding for this work came from UK Science
and Technology Facilities Council Grants No. ST/L000466/1 and No. ST/P000746/1 and Engineering and Physical
Sciences Research Council Project No. EP/W005395/1.

\begin{appendix}

\section{Reconstructing our Results}
\label{fitresrecon}
In the supplementary file \textbf{HPQCD\_BcJpsi\_FF.pydat} we provide 4 synthetic data points for each form factor, at $q^2$ values spanning the full kinematical range and with correlations. These may be loaded using the \textbf{gvar} Python package, and we provide an example script for doing so. We also provide the fit posteriors describing our continuum results in the file \textbf{continuum\_fit\_posteriors.pydat}, as well as python scripts \textbf{CC\_extrapolation.py}, \textbf{CC\_extrapolation\_utilities.py}, \textbf{CC\_fit\_parameters.py}, \textbf{CC\_dispersive\_functions.py} and \textbf{load\_chi\_u12.py} to load these parameters and evaluate the fit function. These scripts have been verified to function correctly with \textbf{Python 3.8.10}~\cite{python3} and packages \textbf{gvar 13.1}~\cite{gvar}, \textbf{numpy 1.24.4}~\cite{numpy} and \textbf{matplotlib 3.7.1}~\cite{matplotlib}.

The matrix elements and masses are included in the supplementary materials file \textbf{matrix\_elements\_and\_masses.pydat} as a Python dictionary intended to be loaded using \textbf{gvar}. The matrix elements are labelled `\textit{Label}\textbf{\_c\_M\_}\textit{$m_h$}\textbf{\_p\_}\textit{$ak$}\textbf{\_}\textit{ensemble}', where `\textit{Label}' corresponds to the labels in~\cref{spintastetable}, \textit{$m_h$} is the heavy quark mass in lattice units, \textit{$ak$} is the momentum in $x$- and $y$-directions and `\textit{ensemble}' labels the set, with labels `flatt', `sflatt', `uflatt', `eflatt', `pflatt' and `psflatt' corresponding to sets 1, 2, 3, 4, 5 and 6 in~\cref{tab:gaugeinfo} respectively. Note that the matrix elements given here include the renormalisation factors, with the tensor renormalisation factors evaluated at scale $\mu=\overline{m}_h(\overline{m}_h)$ as discussed in~\cref{sec:renorm}. Note that, by convention, for the MV matrix elements we multiply by the factor of $ak$ to remove the factor of $1/ak$ multiplied in during correlator fitting, while for MT1 and MA0 we leave this factor in. Functions for converting between our matrix elements and the form factors in the QCD basis are provided in \textbf{CC\_extrapolation\_utilities.py}. The $J/\psi$ and $H_c$ masses are labelled similarly as `\textbf{MJpsi\_}\textit{ensemble}\textbf{c}', and `\textbf{MH\_mh\_}\textit{$m_h$}\textbf{\_}\textit{ensemble}\textbf{c}' respectively.

\section{Correlator Fit Variations}
\label{corrstabsec}

\begin{figure*}
\centering                            
\includegraphics[width=1\textwidth]{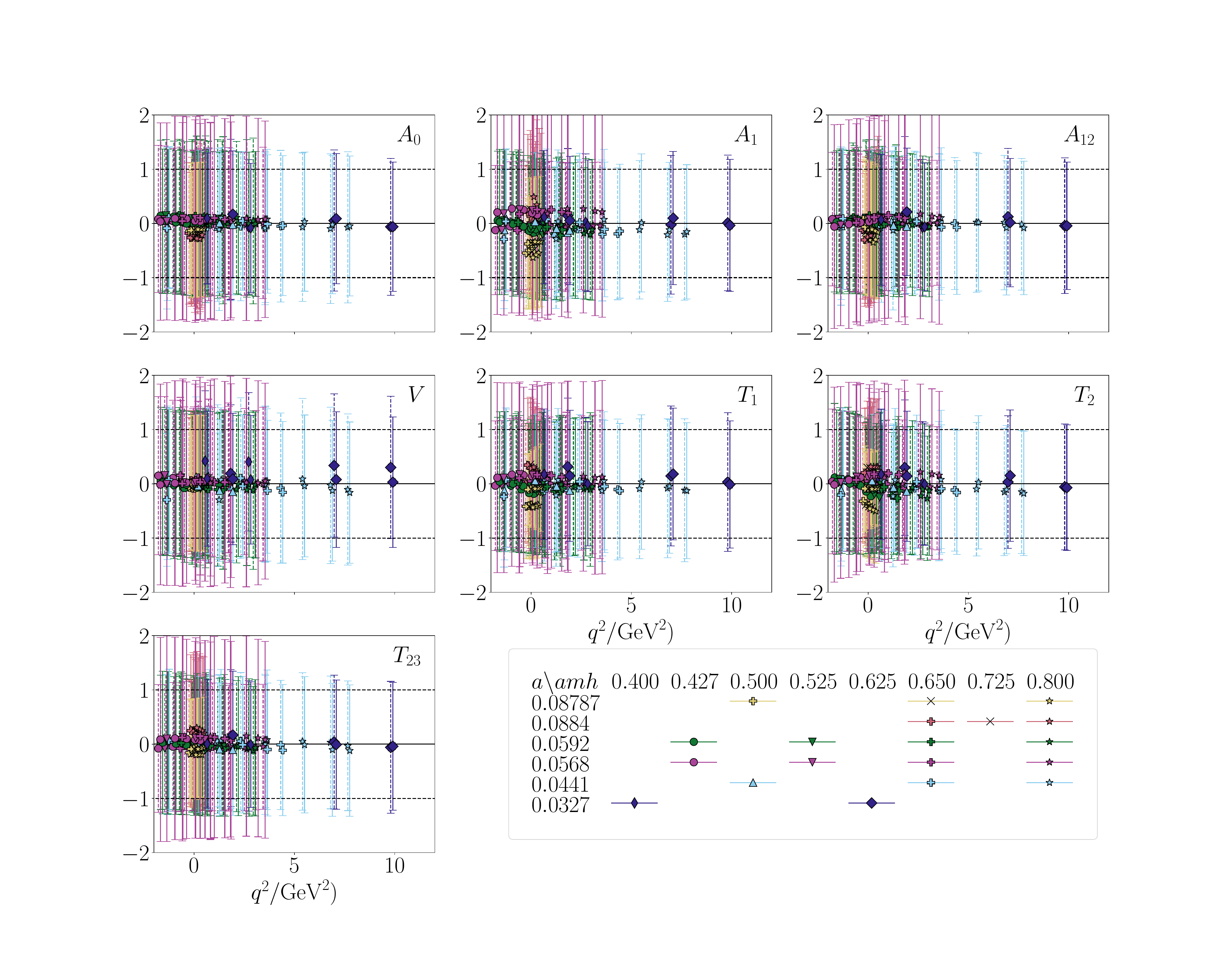}
\caption{\label{ffstabplots}Differences in lattice data points, $(F_Y-F'_Y)/\sigma_{F_Y}$, for the SM and tensor form factors in the QCD basis (defined in~\cref{formfactorsQCDSM,formfactorsQCDNP}) for the different combinations of fit parameters in~\cref{fitparams}, normalised by the standard deviation of the base fit. The data points with dashed(solid) error bars correspond to those fits with $\delta=1(2)$ from~\cref{fitparams}. We see here that the form factors resulting from our correlator fits are essentially unchanged by either increasing the size of the SVD cut or increasing $\Delta T$. Note that we offset the data points slightly in $q^2$ to improve clarity.}
\end{figure*}

\begin{figure}
\centering                            
\includegraphics[width=0.5\textwidth]{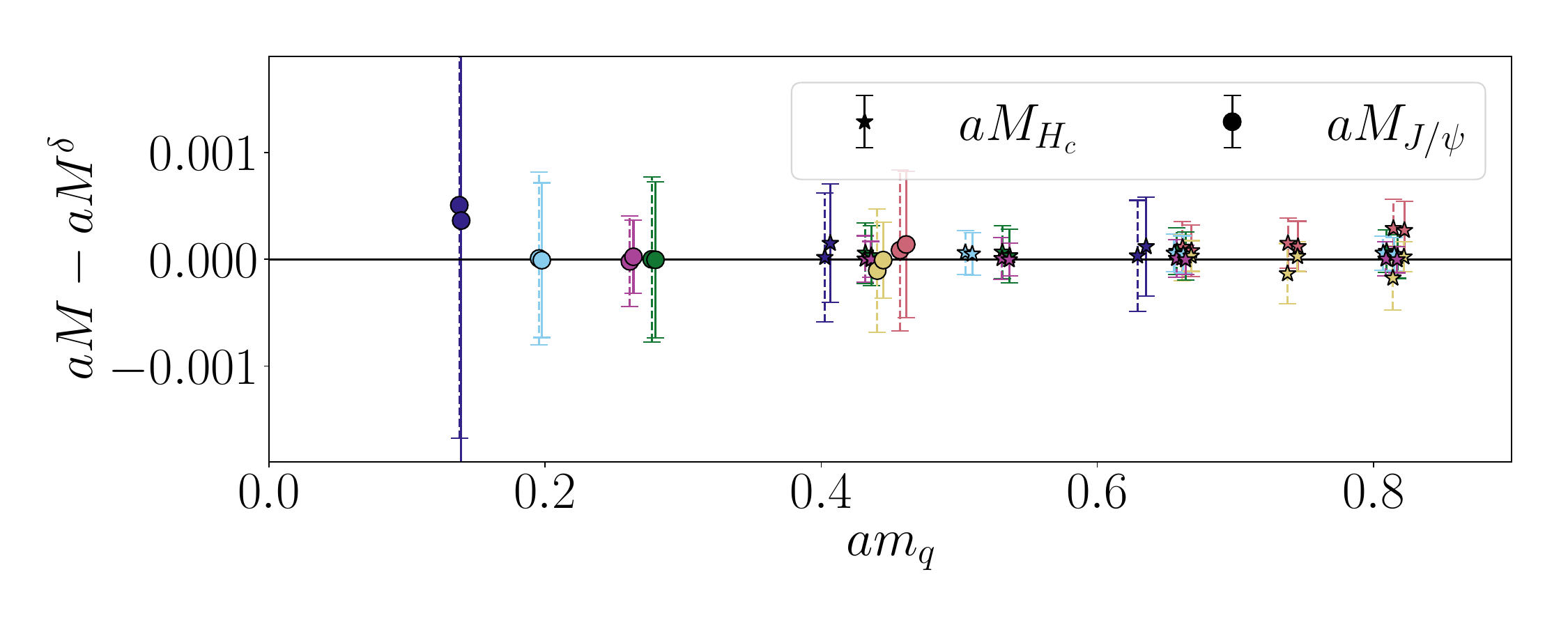}
\caption{\label{Mstabplots} Differences in $J/\psi$ and $H_c$ lattice masses compared to our preferred fit result, $aM$, for the different combinations of fit parameters in~\cref{fitparams}, plotted against either the charm or heavy quark masses in the $J/\psi$ and $H_c$ cases respectively. The data points with dashed(solid) error bars correspond to those fits with $\delta=1(2)$ from~\cref{fitparams}. We see here that the masses resulting from our correlator fits vary by less than $0.05\%$, and by less than $1\sigma$, when increasing the size of the SVD cut or increasing $\Delta T$. The colour scheme used here for the gluon field ensembles is the same as in~\cref{ffstabplots}, and we also offset the data points in $am_q$ to improve clarity.}
\end{figure}

In~\cref{ffstabplots}, we plot the form factors resulting from the different combinations of fit parameters in~\cref{fitparams}. There, we see that the variations of fit parameters result in form factors that are very close to those from the fits chosen for our main analysis. In~\cref{Mstabplots} we plot the changes in $J/\psi$ and $H_c$ lattice masses for the different combinations of fit parameters in~\cref{fitparams}. We see there that the masses resulting from our correlator fits vary by less than $0.1\%$, and by much less than $1\sigma$, when increasing the size of the SVD cut or increasing $\Delta T$. Note also that the values of $\Delta T$ used here are larger than those used in~\cite{Harrison:2020nrv}.

\section{Physical Continuum Fit Function Variations}
\label{contstabsec}

\begin{figure*}
\centering                            
\includegraphics[width=0.95\textwidth]{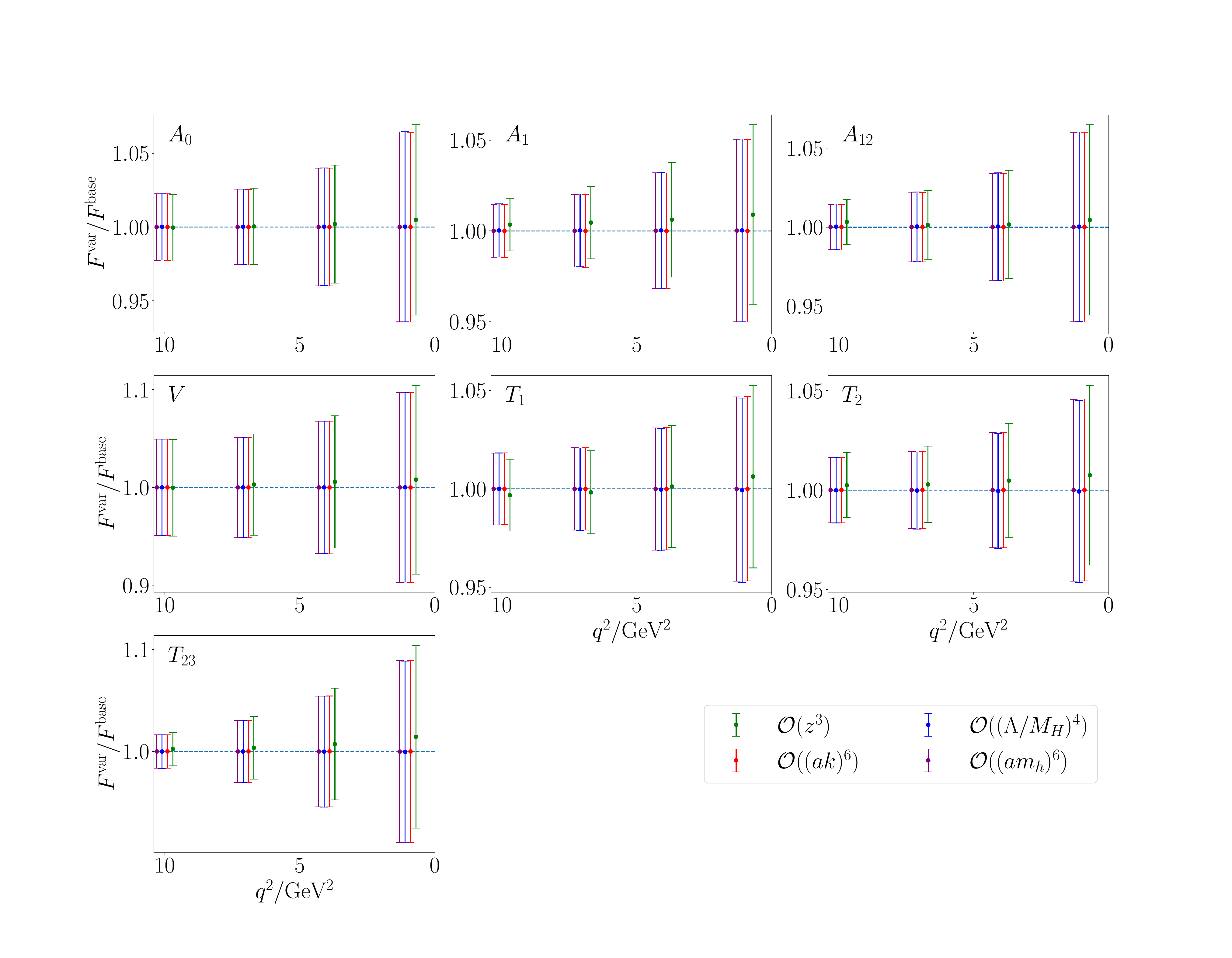}
\caption{\label{fitvarplot} Variations in physical continuum synthetic form factor data points at $q^2=10~\mathrm{GeV}$, $7~\mathrm{GeV}$, $4~\mathrm{GeV}$, and $1~\mathrm{GeV}$ resulting from using only the $\mathcal{O}(z^3)$ orthonormal polynomials, as well as $\mathcal{O}\Big((am_h/\pi)^6\Big)$, $\mathcal{O}\Big((\Lambda/M_H)^3\Big)$, and $\mathcal{O}\Big((ak)^6\Big)$ terms to~\cref{eq:ff}. We show the synthetic data points divided by the values resulting from our main analysis, described in~\cref{physcont}. We see that all variations considered produce results that are consistent well within uncertainties. Note that the plotted values have been offset slightly in $q^2$ to aid readability, with $\mathcal{O}\Big((am_h/\pi)^6\Big)$, $\mathcal{O}\Big((\Lambda/M_H)^3\Big)$, $\mathcal{O}\Big((ak)^6\Big)$, and $\mathcal{O}(z^3)$ fits running from leftmost offset to rightmost respectively.}
\end{figure*}

Here, we investigate the impact of varying our physical continuum fit function, defined in~\cref{eq:ff}, on our results for the $B_c\to J/\psi$ form factors. We consider including one fewer order of orthonormal polynomials, up to $\mathcal{O}(z^3)$, as well as adding $\mathcal{O}\Big((am_h/\pi)^6\Big)$, $\mathcal{O}\Big((\Lambda/M_H)^3\Big)$, and $\mathcal{O}\Big((ak)^6\Big)$. We plot synthetic data at four reference $q^2$ values roughly spanning the full kinematic range resulting from these fit variations in~\cref{fitvarplot}. There, we see that all of the variations considered result in form factors that are consistent with our main result well within $1\sigma$.

\end{appendix}

\bibliographystyle{apsrev4-1}
\bibliography{BcJpsi}

\end{document}